\documentclass{aa}
\usepackage[varg]{txfonts}

\usepackage{xcolor}
\usepackage{siunitx}
\usepackage[inline]{enumitem}
\usepackage{graphicx}
\usepackage{bm}
\usepackage{multirow, array}

\DeclareSIUnit{\pc}{\text{pc}}
\DeclareSIUnit{\dex}{\text{dex}}
\DeclareSIUnit{\Msun}{M_\sun}

\graphicspath{{graphics/}}

\usepackage{natbib}
\usepackage[breaklinks=true]{hyperref}
\bibpunct{(}{)}{;}{a}{}{,} %% natbib format for A&A and ApJ

\newcommand{\OII}{[\ion{O}{ii}]}

\newcommand{\HST}{\textsc{HST}}
\newcommand{\HSTACS}{\textsc{HST-ACS}}

\newcommand{\MUSE}{\textsc{MUSE}}
\newcommand{\MAGIC}{\textsc{MAGIC}}
\newcommand{\COSMOS}{\textsc{Cosmos}}
\newcommand{\COSMOStwenty}{\textsc{Cosmos2020}}

\newcommand{\MUSCATEL}{\textsc{MUSCATEL}}
\newcommand{\MUSEWIDE}{\textsc{MUSE-Wide}}
\newcommand{\HUDF}{\textsc{MUSE-HUDF}}
\newcommand{\HDFS}{\textsc{MUSE-HDFS}}
\newcommand{\Galfit}{\textsc{Galfit}}
\newcommand{\Cigale}{\textsc{Cigale}}

\newcommand{\Mocking}{\textsc{MocKinG}}
\newcommand{\Camel}{\textsc{Camel}}

\newcommand{\matplotlib}{\textsc{matplotlib}}
\newcommand{\scipy}{\textsc{scipy}}
\newcommand{\numpy}{\textsc{numpy}}
\newcommand{\astropy}{\textsc{astropy}}

\newcommand{\NOIISample}{1142}
\newcommand{\NMorphoSample}{890}
\newcommand{\NKinematicSample}{571}
\newcommand{\Nselection}{182}
\newcommand{\NselectionOld}{207}

\begin{document}

	\title{Stellar angular momentum of disk galaxies at $z \approx 0.7$ in the \MAGIC{} survey\thanks{Table \ref{tab:catalog_CDS} is only available in electronic form at the CDS via anonymous ftp to cdsarc.u-strasbg.fr (130.79.128.5) or via http://cdsweb.u-strasbg.fr/cgi-bin/qcat?J/A+A/}\fnmsep\thanks{Based on observations made with ESO telescopes at the Paranal Observatory under programs 094.A-0247, 095.A-0118, 096.A-0596, 097.A-0254, 099.A-0246, 100.A-0607, 101.A-0282, 102.A-0327, and 103.A-0563.}}
	\subtitle{I. Impact of the environment}
	
	\titlerunning{Angular momentum at $z \approx 0.7$ in the \textsc{MAGIC} survey. I. Environment}
	
	\author{W. Mercier\inst{\ref{IRAP}}\fnmsep\thanks{\email{wilfried.mercier@lam.fr}}\and 
	        B. Epinat\inst{\ref{LAM}, \ref{CFHT}}\and 
	        T. Contini\inst{\ref{IRAP}}\and
	        D. Krajnovi\'{c}\inst{\ref{Postdam}}\and 
	        L. Ciesla\inst{\ref{LAM}}\and
	        B. C. Lemaux\inst{\ref{Gemini}}
	        V. Abril-Melgarejo\inst{\ref{STScI}}\and
	        L. Boogaard\inst{\ref{MPIA}}\and
	        D. Pelliccia\inst{\ref{UCO}}
	        }

	\institute{Institut de Recherche en Astrophysique et Planétologie (IRAP), Université de Toulouse, CNRS, UPS, CNES, 31400 Toulouse, France\label{IRAP}
	\and
	Aix Marseille Univ, CNRS, CNES, LAM, Marseille, France\label{LAM}
	\and
	Canada-France-Hawaii Telescope, CNRS, 96743 Kamuela, Hawaii, USA\label{CFHT}
	\and
	Gemini Observatory, NSF’s NOIRLab, 670 N. A’ohoku Place, Hilo, Hawaii, 96720, USA\label{Gemini}
	\and
	Space Telescope Science Institute 3700 San Martin Drive Baltimore, MD 21218, USA\label{STScI}
	\and
	Leibniz-Institut für Astrophysik Potsdam (AIP), An der Sternwarte 16, D-14482 Potsdam, Germany\label{Postdam}
	\and
	Max Planck Institute for Astronomy, Königstuhl 17, D-69117 Heidelberg, Germany\label{MPIA}
	\and
	UCO/Lick Observatory, Department of Astronomy \& Astrophysics, UC Santa Cruz, 1156 High Street, Santa Cruz, CA 95064, USA\label{UCO}
	}

	\abstract{}{%
	At intermediate redshift, galaxy groups/clusters are thought to impact galaxy properties such as their angular momentum. We investigate whether the environment has an impact on the galaxies' stellar angular momentum and identify underlying driving physical mechanisms.}{We derive robust estimates of the stellar angular momentum using Hubble Space Telescope (\HST{}) images combined with spatially resolved ionised gas kinematics from the Multi-Unit Spectroscopic Explorer (\MUSE{}) for a sample of $\sim 200$ galaxies in groups and in the field at $z \sim 0.7$ drawn from the \textsc{MAGIC} survey. Using various environmental tracers, we study the position of the galaxies in the the angular momentum-stellar mass (Fall) relation as a function of environment.}{We measure a \SI{0.12}{dex} ($2\sigma$ significant) depletion of stellar angular momentum for low-mass galaxies ($M_\star < \SI{e10}{\Msun}$) located in groups with respect to the field. Massive galaxies located in dense environments have less angular momentum than expected from the low-mass Fall relation but, without a comparable field sample, we cannot infer whether this effect is mass- or environmentally-driven. Furthermore, these massive galaxies are found in the central parts of the structures and have low systemic velocities.
	
	The observed depletion of angular momentum at low stellar mass does not appear linked with the strength of the over-density around the galaxies but it is strongly correlated with (i) the systemic velocity of the galaxies normalised by the dispersion of their host group and (ii) their ionised gas velocity dispersion.}{%
	Galaxies in groups appear depleted in angular momentum, especially at low stellar mass. Our results suggest that this depletion might be induced by physical mechanisms that scale with the systemic velocity of the galaxies (e.g. stripping or merging) and that such mechanism might be responsible for enhancing the velocity dispersion of the gas as galaxies lose angular momentum.}
	
	\date{Received ; accepted }

	\maketitle
	
	\section{Introduction}
	
	In the current paradigm of galaxy evolution, galaxies are expected to form through the condensation of baryons in the centres of dark matter (DM) haloes where, because of external tidal torques, the gas in the proto-galaxy acquires angular momentum before condensing into a disk and forming stars \citep[e.g.][]{Peebles1969, Fall1980}. The same is true for DM haloes that also acquire angular momentum during their linear phase of structure growth \citep[e.g.][]{Doroshkevich1970, White1984}. Initially, it was thought that the angular momentum of the baryons (e.g. stars or gas) traces that of the DM component. However, recent simulations have shown that this picture is not entirely correct as processes exist that can either add or remove angular momentum from either component independently of the halo and proto-galaxy early formation phase \citep[e.g.][]{Genel2015}. For instance, there is strong evidence supporting galaxies having smoothly accreted large amounts of cold gas from their circum-galactic medium to sustain high values of star formation rate (SFR) across cosmic time \citep[e.g.][]{Bouche2013, Bouche2016, Zabl2019}. This accretion of fresh gas is thought to take place predominantly in the disk plane of late-type galaxies and, thus, does not only drive their star forming events throughout their formation history but also increases the angular momentum associated with their baryonic component \citep[e.g.][]{Danovich2015, Cadiou2022}. Similarly, feedback processes such as galactic winds have also been found as potential mechanisms to increase the angular momentum of the baryons in a galaxy \citep[e.g.][]{DeFelippis2017}. In addition, galaxy mergers can also substantially redistribute angular momentum by increasing that of the DM halo \citep[e.g.][]{Hetznecker2006} and by redistributing the mass, thus, either decrease or increase the angular momentum of the baryons depending on the spins and mass ratio of the two galaxies \citep[e.g.][]{Bois2011, Genel2015, Lagos2018}
	
	Numerous studies have tried to constrain the angular momentum of the stellar and gas components in galaxies at various redshifts since the seminal work of \citet{Fall1983} by studying the angular momentum-stellar mass relation, also known as the Fall relation. In the local Universe, some studies have focussed on the shape of the relation as a function of galaxy type \citep[e.g.][]{Fall1983, Romanowsky2012, Cortese2016, Rizzo2018}, surface brightness \citep[e.g.][]{Salinas2021}, stellar mass \citep[e.g.][]{posti_angular_2018, diteodoro_rotation_2021}, cold gas fraction \citep[e.g.][]{Mancera2021, Kurapati2021}, or bulge fraction \citep[e.g.][]{Obreschkow2014, Fall2018}. The general conclusion that can be drawn is that, in the local Universe, galaxies exhibit a linear relation (in log-log space) between stellar mass and angular momentum otherwise known as the Fall relation (i.e. no deviation at low or high stellar masses). The scatter in the relation seems to be mainly correlated with the galaxies' morphology, with elliptical and disk galaxies with prominent bulges located below the relation found for bulge-less disks. The Fall relation has also been studied at higher redshift \citep[$z\sim 1-2$, e.g.][]{Burkert2016, Swinbank2017, Harrison2017, bouche_muse_2021} where it is found to hold but with a lower zero-point that is consistent with galaxies gaining angular momentum at fixed stellar mass with cosmic time (i.e. with decreasing redshift). Interestingly, the recent analysis by \citet{bouche_muse_2021} hinted at a correlation between the zero-point of the Fall relation and the dynamical state of galaxies in the sense that dispersion dominated systems have lower SFR-weighted angular momentum than rotationally supported galaxies at fixed stellar mass. This correlation was already observed in a few previous studies \citep[e.g.][]{contini_deep_2016, Burkert2016} and it helps to explain the separation in the Fall relation between spirals and ellipticals seen at $z = 0$ \citep[e.g.][]{Romanowsky2012}.
	
	From a methodological perspective, measuring the angular momentum in observed galaxies is not straightforward as it would require to know the full mass distribution (including stars, gas, and DM) as well as the amplitude and orientation of the velocity vector of the component of interest (e.g. stars for the stellar angular momentum) at each position in the galaxy. Hence, given the impossibility to measure the intrinsic gas or stellar angular momentum of the galaxies, proxies have been used in the literature that rely on various assumptions. A common proxy is to assume that the gas and the stars are located in a disk that is dynamically stable against its own gravity through its rotation. With a few additional assumptions on the stellar or gas kinematics and on the mass distribution of the stellar and gas disks, it is possible to derive the \citet[][hereafter RF12]{Romanowsky2012} approximation that is widely used in the literature \citep[e.g.][]{Romanowsky2012, contini_deep_2016, Burkert2016, Swinbank2017, Rizzo2018}. Though already mentioned in \citet{Romanowsky2012} and in subsequent studies, a recent discussion was given in \citet{bouche_muse_2021} where it was shown that, depending on the radius at which the angular momentum is measured, the RF12 approximation can overestimate the galaxies' stellar or gas angular momentum by nearly 20\%. Furthermore, thanks to the increasing number of robust kinematic measurements for the last ten years, some authors have started using other estimates that rely on fewer assumptions. For instance, some studies \citep[e.g.][]{posti_angular_2018, bouche_muse_2021, mancera_pina_baryonic_2021, Mancera2021} assumed an axisymmetric disk model (typically exponential) and either numerically integrated the analytical expression of the angular momentum for a given rotation curve model or substituted the integral with a sum on the pixels of a galaxy's image \citep[e.g.][]{Rizzo2018}. Similarly, \cite{Cortese2016} proposed another method that numerically integrates the angular momentum by summing the contribution of each spatial pixel (spaxel) in a data cube. The main advantage is that it alleviates the assumption of axial symmetry since the observed flux distribution is directly taken into account. However, its main drawback is that it suffers from the usually poor spatial resolution of 3D spectroscopic observations compared to high spatial resolution images obtained, for instance, by the Hubble Space Telescope (\HST{}). Other authors have also implemented more complex methods that combine semi-analytical models with N-body simulations to derive the angular momentum of both the baryonic and DM components of local galaxies \citep[][]{Ansar2023}.
	
	Finally, a few studies have tried to probe the impact of the environment on the galaxies' angular momentum \citep[e.g.][]{pelliccia_searching_2019, Perez2021}. The current picture that emerges from these studies is that galaxies found in galaxy groups and galaxy clusters at $z \sim 1$ seem to have a deficit of angular momentum with respect to field galaxies located at the same redshift. Because physical mechanisms can affect the angular momentum of the baryons in different ways (i.e. increase or decrease it), an interpretation given in \citet{pelliccia_searching_2019} is that this reduction of angular momentum at fixed stellar mass could be due to galaxy mergers, in line with their prevalence in dense environments \citep[e.g.][]{Tomczak2019} and with recent simulations \citep[e.g.][]{Lagos2018}. However, these results required to compare galaxies from different surveys and observed with various instruments \citep[e.g.][]{Perez2021} to reach these conclusions. As in \citet{Abril-Melgarejo2021} and \citet{mercier_scaling_2022}, we argue that there might be systematic effects when doing so, mainly driven by different selection functions between different datasets. 
	
	This paper is the first of a series. In this analysis, we propose to study the impact of the environment on the stellar angular momentum of galaxies in the \MUSE{}-gAlaxy Groups In \COSMOS{} (\MAGIC{}) survey (Epinat et al., in prep.). As shown in \citet{mercier_scaling_2022}, this survey is ideal to probe the impact of the environment on galaxy dynamics at $z \sim 1$ because it allows for the simultaneous observation of galaxies located in structures of various masses (mostly galaxy groups) and foreground/background galaxies in more rarefied environments, as well as to analyse them in a consistent manner. In the second paper of the series, we will investigate how different physical mechanisms can affect galaxies across the Fall relation as a function of their stellar mass, morphology, and dynamical properties.
	
	The structure of the paper is as follows. In Sect.\,\ref{sec:sample}, we give a brief description of the \MAGIC{} sample, the \HST{} and \MUSE{} observations, how we performed the morphological and kinematics modellings, and how we characterised the galaxies' environment. More importantly, we also describe the sample selection used to study the stellar angular momentum. In Sect.\,\ref{sec:angular momentum}, we describe the method used to derive the angular momentum with \HST{} images, and in Sect.\,\ref{sec:Analysis} we assess the reliability of the method. Afterwards, we perform the analysis of the Fall relation in Sect.\,\ref{sec:results} and we conclude in Sect.\,\ref{sec:conclusions}. Throughout this paper, we assume a $\Lambda$ cold dark matter cosmology with ${\rm{H}}_0 = \SI{70}{\kilo\meter\per\second\per\mega\pc}$, $\Omega_{\rm{M}} = 0.3$, and $\Omega_\Lambda = 0.7$.
	
	\section{Sample selection and main properties}
	\label{sec:sample}	
	
	The sample used in this analysis is part of the \MAGIC{} survey (Epinat et al., in prep.), a deep Multi-Unit Spectroscopic Explorer (\MUSE{}) Guaranteed Time Observation (GTO) survey targetting 14 groups in the Cosmic Evolution Survey (\COSMOS{}) area \citep{Scoville2007} with 17 different \MUSE{} pointings. The main goal of this survey is to study the impact of the environment on the properties of galaxies at intermediate redshift by combining multi-band photometry, in particular Hubble Space Telescope Advanced Camera for Surveys (\HSTACS{}) observations \citep{Koekemoer2007, Massey2010}, with spatially resolved spectroscopic properties from \MUSE{}. The analysis performed in this paper is the continuation of two previous ones. In the first paper \citep{Abril-Melgarejo2021}, we studied the impact of the galaxies' environment on the Tully-Fisher Relation (TFR) for a subsample of galaxies located in dense groups. In the second paper  \citep{mercier_scaling_2022}, we focussed our analysis on the impact of the galaxies' environment on three major galaxy scaling relations (size-mass, Main Sequence - MS, and TFR) for the full \MAGIC{} sample. This latter work was carried out by comparing galaxies located in the field with galaxies found in groups with a large dynamical range of density, all observed in \MAGIC{}. A summary of the survey is presented below and for a complete description, see Epinat et al. (in prep.).
	
	\subsection{Observations and physical parameters}
	\label{sec:sample/observations}
	
	Observations are split in 17 different \MUSE{} pointings, resulting in data and variance cubes that range from \SI{4750}{\angstrom} to \SI{9350}{\angstrom} with a spatial sampling of \SI{0.2}{\arcsecond} and a spectral sampling of \SI{1.25}{\angstrom}. The observing strategy and data reduction can be found in Epinat et al. (in prep., but see also \citealp{mercier_scaling_2022} and \citealp{Abril-Melgarejo2021}). The \MUSE{} Line Spread Function (LSF) was modelled with a second order polynomial function as in \citet{bacon_muse_2017} and \citet{guerou_muse_2017} and the \MUSE{} Point Spread Function (PSF) was modelled by extracting \SI{100}{\angstrom} wide narrow-band images of stars in each \MUSE{} field and by fitting them with a Moffat profile with free $\beta$ index. The wavelength dependence of the PSF Full Width at Half Maximum (FWHM) was then derived field-by-field by fitting a linear relation to the median curve (FWHM vs. wavelength) and the $\beta$ parameter as the mean value weighted by the uncertainties. The median value of the \MUSE{} PSF FWHM for the 17 fields is \SI{0.67}{\arcsecond} at \SI{4000}{\angstrom} and \SI{0.53}{\arcsecond} at \SI{8000}{\angstrom} which correspond respectively to \SI{4.8}{\kilo\pc} and \SI{3.8}{\kilo\pc} at $z = 0.7$. In addition, we also used high-resolution $\SI{4}{\arcsecond} \times \SI{4}{\arcsecond}$ \HST{} stamps in the F814W filter\footnote{This is the best \HST{} spatial resolution available in the \COSMOS{} field (PSF FWHM below \SI{0.1}{\arcsecond}, pixel scale of \SI{0.03}{\arcsecond}).} to model the galaxies' morphology. The \HST{} PSF profile was measured in \citet{Abril-Melgarejo2021} by fitting a Moffat profile onto 27 non saturated stars found in the \MUSE{} fields and adopting the median values for the PSF parameters. The PSF parameters used for the morphological modelling are FWHM$_{\rm{HST}} = \SI{0.0852}{\arcsecond}$ and $\beta = 1.9$.
	
	Additionally, we also estimated the galaxies' stellar mass and SFR values with Spectral Energy Distribution (SED) fitting using the Code Investigating GALaxy Emission \citep[\Cigale{}, see][]{boquien_cigale_2019} by fixing the redshift of the galaxies to their \MUSE{} spectroscopic redshift and using \COSMOStwenty{} catalogue of \citet{COSMOS2020}. We used \citet{bruzual_stellar_2003} single stellar populations with a \citet{salpeter_luminosity_1955} initial mass function (IMF) and a single metallicity value of \SI{0.02}{\dex}, a truncated delayed exponential star formation history \citep[SFH, described in][]{ciesla_identification_2018, ciesla_investigating_2021}, and a \citet{charlot_simple_2000} attenuation law with a total-to-selective extinction ratio $R_V = 3.1$. More details about the grid of parameters and the choice of models can be found in Epinat et al. (in prep.). As an indication, we note that different IMF \citep{Chabrier2003} and SFH (exponentially declining) models were used in \citet{Abril-Melgarejo2021} and \citet{mercier_scaling_2022}. In particular, this affects the stellar mass of the galaxies that will be used when investigating the impact of the environment on the galaxies' angular momentum. A comparison between these different models shows overall consistent values within \SI{0.5}{\dex} for galaxies more massive than \SI{e8}{\Msun}. Furthermore, the latest \Cigale{}-based stellar masses are slightly offset by roughly \SI{0.05}{\dex} from the previous values which can be accounted for by the use of a \citet{salpeter_luminosity_1955} IMF instead of a \citet{Chabrier2003} one, as previously done.
	
%	\begin{figure}[htp]
%		\centering
%		\includegraphics[scale=0.8]{comp_mass}
%		\caption{Comparison between the new stellar mass from \Cigale{} and the previous one from \FAST{} used in \citet{mercier_scaling_2022} for the \OII{} emitters sample. A positive difference means the value from \Cigale{} is larger. Black points represent the kinematics sample (see the end of Sect.\,\ref{sec:sample/dynamical modelling}) and red points correspond to other \OII{} emitters.}
%		\label{fig:sample/observations/comp mass SFR}
%	\end{figure}	
	
%	We show in Fig\,\ref{fig:sample/observations/comp mass SFR} the comparison between the new stellar mass derived with \Cigale{} and the previous one used in \citet{mercier_scaling_2022} and derived with \FAST{} (a positive difference means the value from \Cigale{} is higher). Only \OII{} emitters are shown, with galaxies in the kinematics sample (see Sect.\,\ref{sec:sample/dynamical modelling} below) in black.  Above $M_\star \gtrsim \SI{e8}{M_\odot}$ we find stellar masses consistent within $\SI{0.5}{dex}$ and below \Cigale{} tends to find on average lower stellar masses than \FAST{}.
	
	\subsection{Morpho-dynamical modelling}
	\label{sec:sample/dynamical modelling}
	
	In \citet{mercier_scaling_2022}, we carried out a dynamical modelling of the entire \MAGIC{} survey, focussing on galaxies at $0.2 < z < 1.5$, a redshift range where the \OII{} doublet can theoretically be detected given the wavelength coverage of our \MUSE{} observations. Galaxies for which the \OII{} doublet is indeed detected form what we refer to as the kinematics sample (since we can use their \OII{} doublet to extract their ionised gas kinematics). For a complete description of the morphological and kinematics modellings, see Sect.\,4.1 and 5.1 of \citet{mercier_scaling_2022}. In what follows, a quick summary is presented.
	
	First, we modelled the morphology of \NOIISample{} galaxies in the redshift range $0.2 < z < 1.5$ detected in \MAGIC{}. For each galaxy, we performed a bulge-disk decompositions with \Galfit{} \citep{peng_detailed_2002} using an exponential disk model for the stellar disk and a circular de Vaucouleurs profile for the bulge. From this initial sample, \NMorphoSample{} galaxies could be reliably modelled (see Sect.\,4 of \citealt{mercier_scaling_2022} for more details). For each galaxy, various morphological parameters were derived, including:
	\begin{enumerate*}[label=(\roman*)]
		\item their global effective radius $R_{\textit{eff}}$ (i.e. taking into account the mass distribution of both the disk and bulge components, see Eq.\,6 of \citealp{mercier_scaling_2022}),
		\item the bulge-to-total flux ratio (B/T) evaluated at one global effective radius,
		\item the disk major axis position angle (PA), and
		\item the apparent disk axis ratio $q = b/a$, with $a$ and $b$ defined as the major and minor axes, respectively.
	\end{enumerate*}
	Furthermore, we introduced two expressions to correct the central surface brightness and the observed axis ratio $q$ of the stellar disk for its non-zero thickness, assuming an intrinsic double exponential 3D mass distribution (see Sect.\,4.4 and Appendix\,D.4 of \citealp{mercier_scaling_2022} for more details). In addition, because the rotation velocity that was used for the TFR in \citet{Abril-Melgarejo2021} and \citet{mercier_scaling_2022} was measured at the radius $R_{22} = 2.2 R_{\rm{d}}$ in the plane of the disk, with $R_{\rm{d}}$ defined as the disk scale length, we also introduced a corrected stellar mass ($M_{\star, \rm{corr}}$) so that it is also evaluated at $R_{22}$ (see Sect.\,4.3 of \citealp{mercier_scaling_2022}). The same morphological models as in \citet{mercier_scaling_2022} have been used to derive the stellar angular momentum of the galaxies in what follows, except for 17 galaxies for which the morphology was updated (mainly to take into account the impact of bars and spiral arms on the disk parameters, see Appendix\,\ref{appendix:updates_morpho} for a full description of how the models changed for these galaxies). Then, we extracted the kinematics of the ionised gas component in the galaxies using the \OII{} doublet as kinematics tracer. We fitted in each spaxel the \OII{} doublet using \Camel{}\footnote{\url{https://gitlab.lam.fr/bepinat/CAMEL}} and then we cleaned the maps to remove isolated spaxels and those with large velocity discontinuities with respect to their neighbours (see Sect.\,3.5 of \citealp{Abril-Melgarejo2021} or Sect.\,5.1 of \citealp{mercier_scaling_2022} for a complete description). This led to the removal of 271 galaxies that had no remaining spaxels in their kinematics maps with ${\rm{S/N}} > 5$ in the \OII{} doublet. We note that we did not extract the stellar kinematics in \MAGIC{} that would only be available for a small sub-sample of relatively bright/massive galaxies. Therefore, in what follows, we only use the ionised gas kinematics as a proxy of the stellar kinematics to estimate the stellar angular momentum.
	
	Finally, we fitted the velocity field of the ionised gas component of the galaxies using \Mocking{}\footnote{\url{https://gitlab.lam.fr/bepinat/MocKinG}}. We implemented for the rotation curve a mass modelling approach comprising three velocity components:
	\begin{enumerate*}[label=(\roman*)]
		\item a double exponential stellar disk (constrained from the morphology),
		\item a Hernquist stellar bulge (also constrained from the morphology), and 
		\item an unconstrained Navarro-Frenk-White \citep[NFW][]{navarro_structure_1996} DM halo (see Sect.\,5.1 of \citealp{mercier_scaling_2022} for the details of the models).
	\end{enumerate*}
	Note that we do not have any constraints on the cold gas components (i.e. atomic and molecular) of the galaxies	 in \MAGIC{}. Thus, we did not introduce any additional velocity components for the cold gas, which means that in practice the best-fit NFW profile implicitly includes this component along with the DM halo. We used \Mocking{} to model the ionised gas velocity field of the galaxies using the method of line moments \citep{Epinat2010} assuming the gas is located in a razor-thin disk. During the fitting process we fixed the parameters of the stellar disk and bulge components, as well as the centre position and the disk's inclination in order to remove degeneracies. Thus, the only free parameters are:
	\begin{enumerate*}[label=(\roman*)]
		\item the DM halo parameters (scale length $r_s$ and maximum velocity $V_{\rm{h, max}}$),
		\item the kinematics position angle, and
		\item the systemic redshift $z_s$ of the galaxies.
	\end{enumerate*}
	Once a best-fit velocity field model is found, \Mocking{} also derives a beam-smearing and LSF-corrected ionised gas velocity dispersion map. The gas kinematics of the galaxies in \MAGIC{} were re-modelled with respect to \citet{mercier_scaling_2022} using updated Moffat \MUSE{} PSF profiles in each \MUSE{} field (see Epinat et al., in prep.). We also removed four additional galaxies because they were on the edge of the \MUSE{} field-of-view or had signs of merger in their gas kinematics maps, leading to a kinematics sample of \NKinematicSample{} galaxies.
	
	\subsection{Environment characterisation}
	\label{sec:sample/environment}
	
	The galaxies' environment can be characterised in various ways. For instance, the friends-of-friends (FoF) algorithm used in \citet{knobel_zcosmos_2012} and \citet{iovino_high_2016} is easy to implement but it only uses spectroscopic redshifts while the Voronoi tessellation Monte-Carlo mapping (VMC) technique discussed in \citet{Lemaux2017, Lemaux2022} and \citet{Hung2020, Hung2021} combines both photometric and spectroscopic redshifts. Still, every technique has its limitations as it always remains sensitive to the completeness of redshift measurements and the size of the field-of-view. The details of the environment characterisation, including the determination of the groups and the density estimation can be found in the \MAGIC{} survey paper (Epinat et al., in prep.). In what follows, we provide a quick description. To begin with, we defined the groups the galaxies belong to using an iterative 3D FoF algorithm (as in \citealp{Abril-Melgarejo2021} and \citealp{mercier_scaling_2022}). For groups with more than seven galaxy members, the maximum sky projected separation used was \SI{375}{\kilo\pc} and the maximum line-of-sight velocity separation was \SI{500}{\kilo\meter\per\second}. The location of the centre of the groups, their systemic redshift, and their dispersion $\sigma_V$ are determined from the distribution of their galaxy members, and their radius is determined from $\sigma_V$. In particular, the velocity dispersion of the groups is evaluated using the gapper method (see \citealp{Beers1990}, \citealp{Cucciati2010}, or Epinat et al., in prep. for more details). In this analysis we consider three environmental tracers. 
	
	First, we use the richness (i.e. number of galaxy members as given by the FoF algorithm) of the groups. Because it does not take into account the distribution of the galaxies in the groups, we use a second estimate defined as \citep[e.g.][]{Noble2013, pelliccia_searching_2019}
	
	\begin{equation}
		\eta = \frac{R_{\rm{proj}}}{R_{200}} \times \frac{|\Delta v |}{\sigma_V},
	\end{equation}
	where $R_{\rm{proj}}$ is the projected distance of a galaxy with respect to the centre of the group, $R_{200}$ is the radius where the density of the group is equal to 200 times the critical density of the Universe, and $\Delta v$ is the systemic velocity of a galaxy along the line-of-sight with respect to its host group's redshift. Thus, this tracer does not take into account the distribution of galaxies but only their position in phase-space, providing us with an idea of how dynamically bound the galaxy is with respect to its host group. 
	
	The last estimate directly probes the over-density in the vicinity of the galaxies. It relies on the VMC technique (see \citealp{Lemaux2017, Lemaux2022} and \citealp{Hung2020, Hung2021}) where both photometric and spectroscopic redshifts in \COSMOS{} are used to estimate the density $\Sigma$ in cells of $\SI{75}{\kilo\pc} \times \SI{75}{\kilo\pc}$ and $\pm \SI{3.75}{\mega\pc}$ wide (proper distances, see also Epinat et al., in prep.). The over-density $\delta$ is then estimated as $1+\delta = \Sigma / \Sigma_{\rm{med}}$, where $\Sigma_{\rm{med}}$ is the median density over the entire map in the same redshift slice as is used to compute $\Sigma$. For the analysis carried out in Sect.\,\ref{sec:results}, we use the over-density $\delta$ in order to not be biased by sampling variations from one redshift slice to another.

	\subsection{Sample selection}
	\label{sec:sample/selection}
	
	\begin{figure}
		\centering
		\includegraphics[scale=0.7]{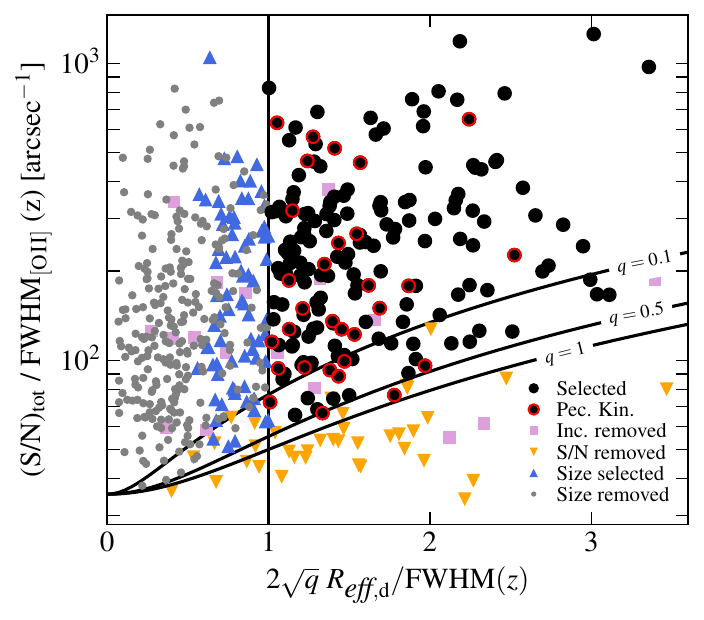}
		\caption{Criteria used for the selection of the kinematics sample. The black points represent galaxies selected according to the surface, S/N, and inclination (removing face-on galaxies only) criteria. Removed galaxies are represented as follows and in this specific order: (i) those removed by the inclination criterion (pink squares), (ii) those removed by the S/N criterion among the remaining galaxies (orange downward pointing triangles), (iii) those removed by the surface criterion among the remaining galaxies but that would have been kept by our previous size criterion used in \citet[][blue upward pointing triangles]{mercier_scaling_2022}, and (iv) those removed by the full selection and by our previous size criterion as well (grey dots). We also show galaxies flagged with peculiar kinematics with red contours. The vertical black line shows the surface selection criterion given in Eq.\,\ref{eq:sample/selection/surface} and the other black lines represent the S/N selection for different disk axis ratios (see Eq.\ref{eq:sample/selection/S/N}).}
		\label{fig:sample/selection/selection}
	\end{figure}
%	
%	In \citet{mercier_scaling_2022} we showed that, depending on the scaling relation studied, different selection criteria had to be used in order to not be biased when fitting these relations. For the analysis of the TFR, we therefore had to remove bulge-dominated, small, and low S/N galaxies. Additionally, we applied two other selections: the first in inclination and the second in stellar mass fraction. Indeed, we had to apply the three first criteria because such galaxies usually have poor constraints on their disk model, especially their half-light radius and axis ratio. Similarly, the inclination criterion was added to remove face-on galaxies whose line-of-sight kinematics is difficult to measure because of projection effects and edge-on galaxies because their mass models are much less constrained. The last selection criterion in \citet{mercier_scaling_2022} was used to select galaxies for which we might have overestimated the contribution of the stellar components to the total rotation curve. These galaxies were visually inspected and those that had consistent morphological and kinematics models (i.e. with an intrinsically high stellar mass fraction) were kept.
	
	Because precisely measuring the stellar angular momentum requires robust estimates of both the morphology and the kinematics of the galaxies, we have updated our sample selection compared to \citet{mercier_scaling_2022}. Since the morphological models were re-inspected for this analysis, instead of applying a B/T criterion to remove galaxies with loosely constrained disks, we manually picked those galaxies and removed them from the sample. In particular, this includes galaxies whose stellar disk component appears noise-dominated in the HST image (as evaluated from the S/N after removing the best-fit bulge component from the image; see Sect.\,5 of \citealp{mercier_scaling_2022} for more details). The five galaxies that were manually removed are \textsc{24\_CGr32}, \textsc{29\_CGr23}, \textsc{240\_CGr30}, \textsc{240\_CGr84}, and \textsc{76\_CGr172}. Because we do not use a B/T criterion, contrary to \citet{mercier_scaling_2022}, we will be able in this analysis to study both disk- and bulge-dominated galaxies. Furthermore, we also decided to apply a selection criterion on the extent of the galaxies on the plane of the sky to make sure that there are enough resolution elements in their \MUSE{} kinematics maps. In this analysis, galaxies are selected if their surface on the plane of the sky within an ellipse with major axis equal to one disk effective radius ($R_{\textit{eff}, \rm{d}}$) and apparent axis ratio ($q$) is larger than the surface of the \MUSE{} PSF within its FWHM. In other words, the size selection writes
	
	\begin{equation}
		2 \sqrt{q} \times R_{\textit{eff}, \rm{d}} / {\rm{FWHM}_{\OII{}}} (z) > 1,
		\label{eq:sample/selection/surface}
	\end{equation}
	where FWHM$_{\OII{}} (z)$ is the \MUSE{} PSF FWHM evaluated at the redshift $z$ of the galaxy and where $R_{\textit{eff}, \rm{d}}$ and FWHM$_{\OII{}}$ must have the same angular or length unit \footnote{We remind that the \OII{} PSF FWHM in Eq.\,\ref{eq:sample/selection/surface} was measured in the \MUSE{} cubes whereas the disk effective radius and apparent axis ratio were both measured in high resolution \HST{} images.}. This criterion naturally excludes galaxies with small sizes with respect to the \MUSE{} PSF and edge-on galaxies because their axis ratio is too small. We also removed face-on galaxies by selecting those with a stellar disk inclination $i > \SI{25}{\degree}$. This value seemed a good compromise between removing galaxies that are too face-on to be well constrained (especially their kinematics) and keeping a sufficiently large sample to perform the analysis. As an indication a more conservative criterion of $i > \SI{30}{\degree}$ would only remove eight additional galaxies. Similarly, we also applied a S/N criterion on the detected \OII{} doublet in the kinematics maps. We selected galaxies if they have an average S/N per spaxel of at least eight across the observed disk's surface on the plane of the sky within the disk's effective radius. Because the PSF smears the stellar disk distribution on the plane of the sky, we assume that the observed disk's extent can be written as the quadratic sum of the intrinsic extent (i.e. $R_{\textit{eff}, \rm{d}}$ and $q R_{\textit{eff}, \rm{d}}$ along the major and minor axes, respectively) and half the PSF FWHM\footnote{In other terms, the observed surface of the stellar disk at an intrinsic distance $R_{\textit{eff}, \rm{d}}$ writes $\pi \sqrt{\left ( R_{\textit{eff}, \rm{d}}^2 + ({\rm{FWHM}}/2)^2 \right ) \left ( q^2 R_{\textit{eff}, \rm{d}}^2 + ({\rm{FWHM}}/2)^2 \right )}$.}. Putting everything together, the S/N selection criterion writes\footnote{The derivation assumes constant flux and S/N maps, as well as Poisson's statistics for the noise.}
	
	\begin{equation}
		\frac{({\rm{S/N}})_{\rm{tot}}}{{\rm{FWHM}}_{\OII{}} (z)} \geq 20 \sqrt{\pi} \left [ \left ( x^2 + 1 \right ) \left (q x^2 + 1 \right ) \right ]^{1/4},
		\label{eq:sample/selection/S/N}
	\end{equation}
	where $x = 2 R_{\textit{eff}, \rm{d}} / {\rm{FWHM}}_{\OII{}} (z)$, with ${\rm{FWHM}}_{\OII{}} (z)$ and $R_{\textit{eff}, \rm{d}}$ in arcsecond. Compared to the previous S/N criterion used in \citet{mercier_scaling_2022} that did not take into account the ellipticity, this one adds 30 new galaxies. However, when including the inclination and surface (size for the old selection) criteria, it removes 25 galaxies. The new sample selection is shown in Fig.\,\ref{fig:sample/selection/selection}. Selected galaxies are shown in black and those removed by the selection with other symbols (one symbol per criterion). Because a galaxy can be removed by different selection criteria, we show them in the following order:
	\begin{enumerate*}[label=(\roman*)]
		\item galaxies removed by the inclination criterion (pink squares),
		\item among the remaining galaxies, those removed by the S/N criterion (i.e. Eq.\,\ref{eq:sample/selection/S/N}; orange downward pointing triangles), and
		\item among the remaining galaxies, those removed by the surface criterion (i.e. Eq.\,\ref{eq:sample/selection/surface}). As an indication, we split the latter between galaxies that would have been selected by the size selection criterion used in \citet{mercier_scaling_2022} (blue upward pointing triangles) and galaxies that that would have been removed by this size selection criterion (grey dots).
	\end{enumerate*}
	The black vertical line shows the limit of the surface selection criterion (see Eq.\,\ref{eq:sample/selection/surface}) and the other black lines show various S/N criteria with different disk axis ratios $q$ (see Eq.\,\ref{eq:sample/selection/S/N}). The combination of the surface, S/N and inclination criteria yields a sample of \Nselection{} galaxies. As an indication, the previous selection used in \citet{mercier_scaling_2022} would have yielded \NselectionOld{} galaxies instead. Among them, those that would have been added by the size criterion are mostly located at the limit of the selection (thus quite small) and are significantly inclined ($i \gtrsim \SI{60}{\degree}$).
	
	Finally, we decided to visually inspect the kinematics maps and the rotation curves of the remaining galaxies and to flag the 30 following galaxies as having 'peculiar kinematics'\footnote{See Table\,F.1 of \citealp{mercier_scaling_2022} or Epinat et al., in prep. for the galaxies' IDs.}:
	\begin{enumerate*}[label=(\roman*)]
		\item \textsc{38\_CGr172}, \textsc{54\_CGr51}, \textsc{70\_CGr79}, \textsc{74\_CGr172}, \textsc{90\_CGr23}, \textsc{93\_CGr114}, \textsc{101\_CGr32}, \textsc{104\_CGr28}, \textsc{104\_CGr172}, \textsc{105\_CGr114}, \textsc{113\_CGr23}, \textsc{148\_CGr30}, \textsc{185\_CGr30}, \textsc{226\_CGr84}, \textsc{257\_CGr84}, \textsc{267\_CGr84}, \textsc{313\_CGr84}, \textsc{442\_CGr32}, and \textsc{454\_CGr32} because there is no visible velocity gradient in their velocity fields so that their best-fit rotation curve can hardly describe the intrinsic rotation of the gas,
		\item \textsc{28\_CGr26}, \textsc{96\_CGr28}, and \textsc{172\_CGr32} because there is no central peak observed in the \HST{} image as would be expected for an exponential disk even without a bulge so that the contribution of their disk component is overestimated in the inner parts,
		\item \textsc{23\_CGr84} because it has a very massive disk whose contribution is overestimated in the inner parts so that its contribution to the total rotation curve is too high and therefore produces large velocity residuals,
		\item \textsc{87\_CGr35} and \textsc{345\_CGr32} because their velocity field and \OII{} emission are quite off-centred from their morphological centre so that their velocity gradient is not correctly fitted in the inner parts,
		\item \textsc{37\_CGr84}, \textsc{85\_CGr23}, and \textsc{278\_CGr84} because, even though they do show signs of rotation, their velocity fields are too perturbed for a rotating razor-thin gas disk model to properly fit their kinematics,
		\item \textsc{106\_CGr84} because there are two kinematically distinct components in its velocity field whose centres match the locations of two morphologically distinct component in its \HST{} image (perhaps following a merging event), therefore rendering the kinematics model uncertain, and
		\item \textsc{85\_CGr35} because it is a double-peaked galaxy in its \HST{} image with little rotation and whose velocity field residuals show that a rotating razor-thin gas disk model is not sufficient.
	\end{enumerate*}
	Rather than removing these galaxies, we will keep them in what follows and we will clearly identify them when analysing their stellar angular momentum.
	
	The choice of whether a galaxy has enough rotation to be flagged or not is ultimately a question of perspective. Thus, we have tried to remain as conservative as possible. To illustrate the difference between a galaxy that does not have clear signs of rotation and one that does, as well as galaxies at the limit of this selection, we show examples of dynamical models for such cases in Fig.\,\ref{fig:sample/selection/examples dynamical models}. For each sub-figure the morphological model is shown in the leftmost column, the velocity field model in the middle one, and the velocity dispersion in the rightmost one. We show on the top left galaxy \textsc{18\_CGr114} whose morphology and kinematics are sufficiently well fitted for the analysis of the angular momentum and on the top right \textsc{70\_CGr79} that does not have any velocity gradient in its velocity field (i.e. flagged). On the bottom row, we show on the left-hand side galaxy \textsc{17\_CGr34} that has a small and slightly disturbed velocity field (not flagged) and on the right-hand side \textsc{278\_CGr84}, a massive galaxy with a large but quite disturbed velocity field as can be seen in its velocity field residual map. Because the rotating razor-thin gas disk model produces large residuals, it is unlikely that we correctly constrain the intrinsic gas kinematics and so we decided to flag it.

	\section{Deriving the stellar angular momentum}
	\label{sec:angular momentum}
	
	We are interested in deriving as precisely as possible the stellar angular momentum of the galaxies. We therefore implement a method, discussed in Sects.\,\ref{sec:angular momentum/general} and \ref{sec:angular momentum/hst maps}, that makes full use of the morphological information contained in the high-resolution \HST{} images and of the best-fit gas kinematics (velocity field and dispersion map) derived from the \MUSE{} data cubes in the previous section. We assume that the ionised gas kinematics traces sufficiently well the stellar kinematics to derive the stellar angular momentum. This assumption is based on comparisons performed between the kinematics of the stellar and ionised gas components of galaxies at intermediate redshift \citep[e.g.][]{guerou_muse_2017} and is discussed in more details in Sect.\,\ref{sec:angular momentum/application}. The formalism presented below only applies for (potentially thick) stellar disks. Because we assume that bulges are spherically symmetric, we therefore cannot use this formalism to estimate their angular momentum. In addition, we expect bulges to be dispersion dominated and therefore to have a negligible contribution to the total stellar angular momentum compared to that of the disk component (see Sect.\,\ref{sec:angular momentum/application} for more details). Thus, in what follows, we focus our analysis on the angular momentum of the stellar disk component only.

	\subsection{General derivation}
	\label{sec:angular momentum/general}
	
	Deriving an accurate estimate of the stellar angular momentum $\vec J_\star$ in galaxies is not straightforward given that it involves knowing \textit{a priori} the 3D stellar mass density $\rho_{\rm{M}}$ as well as the 3D velocity vector $\vec V$ of the stars. In general terms, the stellar angular momentum integrated in a volume $\mathcal{V}$ writes
	
	\begin{equation}
		\vec J_\star (\vec r) = \int_{\mathcal{V}} d^3 \vec r\,\rho_M (\vec r)\,\vec r \times \vec V (\vec r),
		\label{eq:derivation/definition_angular_momentum}
	\end{equation}
	where $\times$ represents the cross product operation. However, Eq.\,\ref{eq:derivation/definition_angular_momentum} is hardly directly usable unless one is working with simulations \citep[e.g.][]{brook_hierarchical_2012, stewart_angular_2013, zavala_link_2016, cadiou_stellar_2022} or in the vicinity of the Milky Way where 6D phase-space positions are available from Gaia \citep[e.g.][]{del_pino_revealing_2021}. Therefore, assumptions must be made on both the stellar mass density and the velocity of the stars in order to constrain their angular momentum from morphological and kinematics data. Perhaps the most widely used expression in the literature is that of RF12 which assumes that the stellar mass distribution can be described by a razor-thin exponential disk with a constant rotation curve. Recently, \citet{posti_angular_2018} and \citet{bouche_muse_2021} showed that this approximation may not correctly compute the stellar angular momentum of intermediate redshift galaxies, especially in the case of low-mass galaxies (typically $\log_{10} M_\star/{\rm{M_\odot}} \lesssim 9-9.5$) since they tend to have a shallower inner slope in their rotation curve than higher-mass counterparts. A more general expression than RF12 can be derived from Eq.\,\ref{eq:derivation/definition_angular_momentum} assuming a razor-thin disk under rotation only, that is neglecting radial and vertical motions (see Appendix\,\ref{appendix:angular momentum} for a derivation), without making any assumptions on the shape of the rotation curve. In this case, the stellar angular momentum becomes orthogonal to the plane of the stellar disk. Furthermore, it is common practice to normalise it by the stellar mass of the galaxy, in which case it is referred to as the specific angular momentum defined as
	
	\begin{equation}
		j_\star (\mathcal{S}) = \int_{\mathcal{S}} dS R~\Sigma_{\rm{M}} (R, \theta)\,V_\theta (R, \theta) / M_\star,
		\label{eq:derivation/specific_angular_momentum_simplified}
	\end{equation}
	where $\mathcal{S}$ is the surface in the plane of the stellar disk over which the angular momentum is integrated, $dS = R d\theta dR$ is the integrand, $R$ is the distance to the centre of the galaxy, $\theta$ is the azimuthal angle in the plane of the disk, $\Sigma_{\rm{M}}$ is the surface mass density of the stellar disk component (i.e. the integral of the 3D mass distribution along the vertical direction with respect to the plane of the disk), and $V_\theta (R, \theta)$ is the rotation velocity of the stars at position $(R, \theta)$. Even though there is no restriction on the surface $\mathcal{S}$ onto which the stellar angular momentum is integrated, it is nevertheless a common practice to integrate it in a circular aperture of radius $R$. If we further assume that both $\Sigma_{\rm{M}}$ and $V$ are independent of $\theta$ and that the mass-to-light ratio $\Upsilon_\star$ is constant throughout the galaxy's stellar disk, Eq.\,\ref{eq:derivation/specific_angular_momentum_simplified} simplifies to
	
	\begin{equation}
		j_\star\,(R) = 2\pi \Upsilon_\star \int_0^{R} dR'\,R'^2~\Sigma (R')\,V_\theta (R') / M_\star,
		\label{eq:derivation/specific_angular_momentum_this_analysis}
	\end{equation}
	where $\Sigma$ is the intrinsic surface brightness distribution (i.e. not sky projected). Equation\,\ref{eq:derivation/specific_angular_momentum_this_analysis} corresponds to the expression used in \citet{bouche_muse_2021} and \citet{mancera_pina_baryonic_2021}. On the other hand, Eq.\,\ref{eq:derivation/specific_angular_momentum_simplified} is slightly more general because it can account for both axisymmetric and non-axisymmetric stellar disks. Both equations are also valid for thick stellar disks\footnote{Equations\,\ref{eq:derivation/specific_angular_momentum_simplified} and \ref{eq:derivation/specific_angular_momentum_this_analysis} are not valid in the case of spherical bulges since it is not possible to write the mass distribution as the product of a surface density and a vertical profile. Therefore, in what follows, we restrict ourselves to estimating the stellar angular momentum of the disk component only.} as long as the disks are 
\begin{enumerate*}[label=(\roman*)]	
	\item symmetric with respect to the plane of the disk and that 
	\item it is possible to separate the vertical profile from the surface brightness distribution in the 3D stellar mass density (see Appendix\,\ref{appendix:angular momentum/thick disks}). 
\end{enumerate*}
The choice of normalisation is technically free. In this analysis, we always normalise the stellar angular momentum using the mass of the stellar disk component integrated in the same radius within which the angular momentum is estimated (i.e. within $R_{22} = 2.2 R_{\rm{d}}$, where $R_{\rm{d}}$ is the stellar disk's scale length, see also Sect.\,\ref{sec:Analysis}).

	\subsection{Estimating angular momentum from \textsc{HST} maps}
	\label{sec:angular momentum/hst maps}
	
	Instead of directly using Eq.\,\ref{eq:derivation/specific_angular_momentum_this_analysis}, we decided to use the more general expression given by Eq.\,\ref{eq:derivation/specific_angular_momentum_simplified} in combination with \HST{} images to take into account the non-axisymmetric mass distribution of the galaxies' stellar disk component. To do so, we approximate Eq.\,\ref{eq:derivation/specific_angular_momentum_simplified} by discretising it along the \HST{} images' spatial dimensions $x'$ and $y'$. We consider $R$, $V_\theta$, and $\Sigma_M$ constant within a pixel's surface and we assume a constant mass-to-light ratio throughout the galaxy's stellar disk so that $\Sigma_{\rm{M}} = \Upsilon_\star \Sigma$, with $\Sigma$ the intrinsic surface brightness distribution of the disk, and $M_\star = \Upsilon F_{\rm{tot}}$, with $F_{\rm{tot}}$ the total flux integrated in the same aperture as $M_\star$. Then, each pixel at position $(x', y')$ contributes to the specific stellar angular momentum as
	
	\begin{equation}
		j_{\star, \rm{pix}} (x', y') = R\,F(x', y') V_\theta (x', y') / F_{\rm{tot}},
		\label{eq:hst maps/specific_angular_momentum_pixel}
	\end{equation}
	where $R = \sqrt{x^2 + y^2}$ is the radial distance at position $(x, y)$ in the plane of the disk and $F$ is the flux in the pixel. The conversion from the apparent position $(x', y')$  on the the sky to its location $(x, y)$ in the plane of the disk is done by taking into account the position of the galaxy's centre, the position angle (PA), and the inclination of the disk, all derived in \citet{mercier_scaling_2022} from the bulge-disk decomposition performed with \Galfit{}, assuming a razor-thin disk geometry (i.e. elliptical isophotes). The specific stellar angular momentum within a circular aperture of radius $r$ in the plane of the stellar disk is then given as the sum of the contribution of each pixel in the aperture, that is
	
	\begin{equation}
		j_{\star} (< r) = \displaystyle\sum_{\lbrace x', y'\,|\,R < r \rbrace} j_{\star, \rm{pix}} (x', y').
        \label{eq:hst maps/specific_angular_momentum_this_analysis}
	\end{equation}
	
	We cannot directly use the ionised gas velocity fields extracted from the \MUSE{} cubes to compute the stellar angular momentum for a few reasons. First, the \MUSE{} observations are much less spatially resolved than the \HST{} data (\SI{0.2}{\arcsecond} per spaxel for a PSF FWHM of \SI{0.5}{\arcsecond} on average for \MUSE{} versus \SI{0.03}{\arcsecond} per pixel for a PSF FWHM of roughly \SI{0.1}{\arcsecond} for \HST{}, see Sect.\,\ref{sec:sample/observations}). Second, the velocity fields extracted from the cubes are too severely affected by beam smearing, especially in the inner parts where the velocity gradient and the ionised gas flux are large. Third, the velocity fields are projected onto the sky and there is no trivial way to invert the projection (especially along the minor axis). Thus, we use instead the rotation curves obtained from the forward mass models performed on the velocity fields in \citet{mercier_scaling_2022}. These have the advantage of being intrinsic (i.e. before the impact of beam-smearing and projection effects) and they can be interpolated to any radius $r$.

	\subsection{Application of the \HST{} formalism to \MAGIC{}}		
	\label{sec:angular momentum/application}
	
	Our goal is to estimate the stellar angular momentum of intermediate redshift galaxies in \MAGIC{}. To do so, we need both the distribution and the velocity of the stars in the plane of the galaxies' stellar disk. The former can be estimated from the HST F814W images, assuming this band traces sufficiently well the stellar mass of the galaxies at $z \approx 1$, but the latter would only be available for a limited number of galaxies. Indeed, our sample is comprised mostly of star-forming galaxies with strong emission lines (e.g. \OII{}) for which we expect only a small fraction (roughly less than 25\%) to have strong enough absorption lines to derive their stellar kinematics. Hence, we have decided to estimate the stellar angular momentum by using the ionised gas rotation curves as proxy for the stellar kinematics. This approximation assumes
\begin{enumerate*}[label=(\roman*)]
	\item that there is co-rotation between the gas and the stars and
	\item that the stars are located in dynamically cold disks (i.e. low stellar velocity dispersion).
\end{enumerate*}
If a galaxy is in equilibrium, then the dynamics of both the stars and the gas should be governed by the galaxy's total gravitational potential and its components should have comparable circular velocities. However, one caveat is that each component might have different velocity dispersions which will contribute to the dynamical support (i.e. asymmetric drift) and will therefore lower the rotation velocities by different amounts. For instance, this is illustrated in \citet{guerou_muse_2017} who showed that even though there are variations between the rotation of the gas and the stars, both components have similar kinematics when taking into account the effect of the velocity dispersion. However, because we do not have any estimates of the stellar velocity dispersion, we cannot correct \textit{a posteriori} for this effect.
	
	We note that similar methods to the one we have implemented have already been used in a few previous studies \citep[e.g.][]{cortese_sami_2016, diteodoro_rotation_2021}. However, they usually take into account the contribution of all the components (in general stellar disk and bulge) to the observed surface brightness distribution. Doing so may not be entirely appropriate because
	\begin{enumerate*}[label=(\roman*)]
		\item the equations that are used are only valid in the case of a disk, and
		 \item bulges might be dispersion dominated systems in which case their angular momentum should be nearly null.
	\end{enumerate*}
	Thus, to avoid being biased by the bulge component that can significantly contribute to the flux in the inner parts, we removed the best-fit \Galfit{} bulge model from the \HST{} images before computing the stellar angular momentum. In the rare cases where the flux in the bulge-removed images becomes negative near the centre (e.g. because its contribution was slightly overestimated by \Galfit{}), we replace the pixels with negative values by the flux of the disk model at the same location. Because we use in Eq.\,\ref{eq:hst maps/specific_angular_momentum_pixel} the flux of the pixels, we know that the estimate of the stellar angular momentum is also going to be impacted by the noise in the \HST{} images. Given that we had already removed the background signal from the images, either beforehand or by using an additional sky background component during the morphological modelling performed with \Galfit{}, we expect the noise to have a null mean. Hence, if there is a sufficiently large number of pixels in the sum in Eq.\,\ref{eq:hst maps/specific_angular_momentum_this_analysis}, then the contribution of the noise to the stellar angular momentum should be close to null. This argument does not hold near the centre where there are not enough pixels to properly sample the noise distribution. However, since the rotation velocity also quickly drops to zero there, the impact of noise should also be reduced in the inner parts. 
	
	Finally, we note that the we could have also used the best-fit \Galfit{} disk model instead of relying on bulge-removed \HST{} images to derive the stellar angular momentum of the galaxies. This would alleviate the difficulties of dealing with projection effects and the impact of noise but the main drawback in doing so is that we would not be able to take into account asymmetries in the disks of the galaxies since our model assumes axial symmetry. In Sect.\,\ref{sec:Analysis/reliability}, we perform a comparison between the angular momentum derived from \HST{} images and that derived from the best-fit \Galfit{} disk model but, in the following parts of Sect.\,\ref{sec:Analysis}, we solely rely on bulge-removed \HST{} images to estimate the angular momentum.
	
	\section{Analysis}	
	\label{sec:Analysis}
	
	We present in this section the framework used to analyse the impact of the environment on the stellar angular momentum of the galaxies. First, we discuss in Sect.\,\ref{sec:Analysis/Methodology} the methodology used to fit the Fall relation and we discuss the effect of the selection on the shape of the Fall relation for the entire kinematics sample. Then, we asses in Sect.\,\ref{sec:Analysis/reliability} the reliability of the \HST{} formalism used in this work by comparing it with other angular momentum estimates.
	
	\subsection{Methodology}
	\label{sec:Analysis/Methodology}
	
	We use the \Nselection{} galaxies from the kinematics sample, 30 of which have been flagged in Sect.\,\ref{sec:sample/selection} as having peculiar kinematics. In what follows, we consider the formalism based on \HST{} images discussed in Sect.\,\ref{sec:angular momentum/hst maps}. In this analysis, the angular momentum (including its normalisation taken as the mass of the disk component only) and the stellar mass appearing in the Fall relation are always measured within $R_{22}$\footnote{We note that photometric and kinematics measurements beyond this radius become less robust because of the significant drop in S/N.}. Similarly, we always remove the contribution of the bulge component to the \HST{} images before estimating the galaxies' stellar angular momentum. 
	
\begin{figure}[htp]
		\centering
		\includegraphics[scale=0.9]{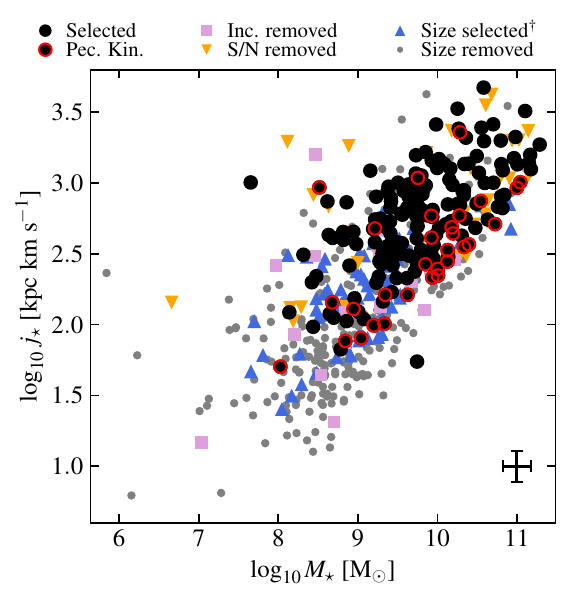}
		\caption{Fall relation for the entire kinematics sample, using bulge-removed \HST{} images. Galaxies selected from the kinematics sample are represented with black points and other symbols represent galaxies removed by different selection criteria (see Fig.\,\ref{fig:sample/selection/selection}). Red contours correspond to galaxies flagged with peculiar kinematics. The typical uncertainty (including systematic uncertainties added when fitting the relation) is shown as an error bar on the bottom right corner. $^\dagger$Galaxies removed by the surface criterion defined in Sect.\,\ref{sec:sample/selection} but that would have been kept by the size selection criterion used in \citet{mercier_scaling_2022}.}
		\label{fig:analysis/momentum first plot}
	\end{figure}

	The Fall relation for the entire kinematics sample is shown in Fig.\,\ref{fig:analysis/momentum first plot}, where we use symbols similar to those in Fig.\,\ref{fig:sample/selection/selection}. In particular, selected galaxies are represented with black circles and those flagged as having peculiar kinematics are shown with red contours. We recover the usual shape of the Fall relation with the angular momentum that seems to scale roughly linearly with stellar mass. Our range of angular momentum values and the scatter in the relation are also consistent with recent results obtained with \MUSE{} using a SFR-weighted angular momentum tracer \citep{bouche_muse_2021}. There is no significant difference in the shape of the relation between galaxies that are kept by the selection (i.e. black circles) and those that are removed (i.e. other symbols). Low-mass objects ($M_\star \lesssim \SI{e9}{\Msun}$) mostly correspond to small galaxies, as would be expected from the shape of the size-mass relation \citep[see Fig.\,10 of][]{mercier_scaling_2022}. Incidently, this means that very few galaxies are kept below this limit (18 in total versus 152 for the galaxies removed by the selection) and that this analysis will therefore focus on galaxies in the stellar mass range $M_\star \sim 10^9 - \SI{e11}{\Msun}$. 
	
	Another important point is that flagged galaxies mostly populate the bottom part of the relation. This can be understood by noting that the majority of these galaxies do not exhibit visible velocity gradients, meaning that their rotation velocity is poorly constrained. But, because the contribution to the rotation curve of the stellar disk and bulge components is entirely fixed by the morphology, applying a mass modelling approach for these galaxies can result in a velocity field model that rotates faster than the observed one. In such cases, no DM halo can be fitted because the rotation velocity is already too high with just the bulge and disk components. The net effect is that it will produce a spurious correlation between the stellar mass and angular momentum of the galaxies (described by Eq.\,\ref{eq:appendix/angular momentum/special/disk/R22} in the case of a bulge-less galaxy; see also the end of Sect.\,6 of \citealt{mercier_scaling_2022} for a discussion of the limits of mass models for such cases). One way to alleviate these spurious correlations is to use for comparison a non-physically motivated rotation curve model (i.e. a model that does not use the morphology as a prior or constraint for the kinematics). Thus, in what follows, we will always analyse our results using both the mass model rotation curve, as well as a flat rotation curve model (composed of a linearly rising part followed by a plateau, see also Eq.\,7 of \citealt{Abril-Melgarejo2021}).
	
	To study the impact of the galaxies' environment on the shape of the Fall relation, we use the same methodology as applied to the TFR in \citet{mercier_scaling_2022}, that is we fit
	
	\begin{equation}
		\log_{10} j_\star\,[\unit{\kilo\pc\kilo\meter\per\second}] = \beta + \alpha (\log_{10} M_\star \,[\unit{\Msun}] - p),
	\end{equation}
	where $\beta$ is the the zero-point, $\alpha$ is the slope, and $p = \SI{9.8}{\dex}$ is the pivot point taken as the decimal logarithm of the median stellar mass (in solar masses) of the kinematics sample\footnote{We include a pivot point in order to reduce the correlation between the slope and the zero point during the fit.}. We always consider $\log_{10} j_\star$ as the dependent variable and $\log_{10} M_\star$ as the independent variable. Following \citet{mercier_scaling_2022}, we will mostly investigate variations in the zero-point of the Fall relation as a function of galaxy environment for a fixed slope. Nevertheless, we will also quickly discuss a potential variation of the slope of the relation, especially at the high-mass end, though a more thorough discussion will be given in the second paper of the series. We use a slightly updated version of \textsc{LtsFit}\footnote{\url{https://pypi.org/project/ltsfit/}} \citep{Cappellari2013a} to fit the relation which implements the Least Trimmed Squares (LTS) technique (see \citealp{Rousseeuw2006}) to find and remove outliers from the fit. The only feature that this new version adds is the possibility to fit with a fixed slope. When doing so, we always use the following procedure:
	\begin{enumerate*}[label=(\roman*)]
		\item we fit the entire sample with a free slope, and
		\item we fix the slope to this value before fitting each subsample separately.
	\end{enumerate*}
	When fitting the Fall relation, we take into account the uncertainties on both dependent and independent variables.	For the stellar mass, we use the uncertainty derived from SED fitting and we add a systematic uncertainty of \SI{0.2}{\dex}, as in \citet{mercier_scaling_2022}. For each galaxy, we estimate the uncertainty on $j_\star$ by generating 100 Monte-Carlo realisations of the flux distribution and kinematics parameters of the rotation curve. For each realisation, the stellar angular momentum is estimated and the uncertainty on $j_\star$ is taken as the standard deviation of the 100 realisations. Furthermore, in order not to treat asymmetrically the dependent and independent variables (i.e. having one variable with underestimated uncertainties) during the fitting process, we also add a systematic uncertainty of \SI{0.1}{\dex} to the angular momentum. Finally, we estimate for each fit the unbiased average, standard-deviation, and 95\% confidence interval (CI) of the best-fit zero-point using jackknife resampling \citep{QUENOUILLE1956, Tukey1958} on the considered sub-sample with the slope and outliers fixed.
	
	\subsection{Assessing the reliability of the method}
	\label{sec:Analysis/reliability}
	
	\begin{figure*}[htp]
		\centering
		\includegraphics[scale=0.9]{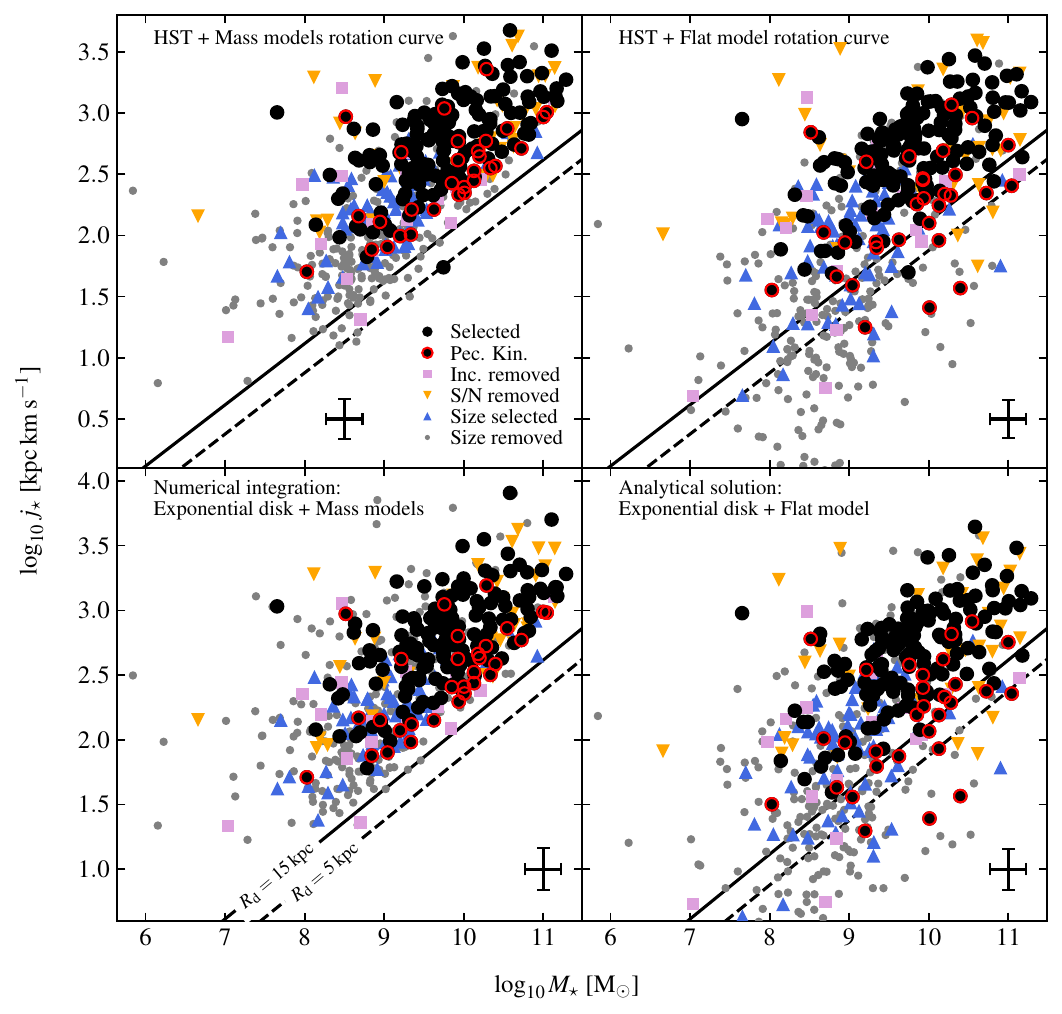}
		\caption{Fall relation for the kinematics sample using various formalisms. On the top row is shown the angular momentum derived using \HST{} maps and on the bottom row that derived using an exponential disk model. The leftmost column represents the Fall relation when using the rotation curve derived from the mass modelling and the rightmost column the Fall relation when using a flat model rotation curve (the top-left plot is similar to Fig.\,\ref{fig:analysis/momentum first plot}). Galaxies selected for the analysis are shown as black points and those flagged with peculiar kinematics with red contours. Other symbols correspond to galaxies removed by the selection (see Fig.\,\ref{fig:sample/selection/selection} for more details). The typical uncertainty is shown on the bottom of each plot as a black error-bar. The two black lines show the limit of the Fall relation for an exponential disk without DM halo and with two different disk scale lengths of \SI{5}{\kilo\pc} (dashed line) and \SI{15}{\kilo\pc} (plain line, see Eq.\,\ref{eq:appendix/angular momentum/special/disk/R22}).}
		\label{fig:analysis/reliability/Fall relation HST comparison}
	\end{figure*}
	
	In this section, we estimate the reliability of the method we used to derive the stellar angular momentum. To do so, we can first check the impact of using \HST{} images instead of assuming a stellar mass distribution that follows an axisymmetric disk. Then, we can see how changing the shape of the rotation curve affects the value of the stellar angular momentum. For that, we can compare its value when using the mass model rotation curve (i.e. including the contributions of the stellar bulge, stellar disk, and DM halo) with the value obtained when using a simpler flat model rotation curve (see \citealp{Abril-Melgarejo2021} for a definition). Even though the latter is not physically motivated, it is nevertheless a robust model that was fitted in \citet{mercier_scaling_2022} to the ionised gas kinematics to asses the reliability of the recovered gas intrinsic velocities. The Fall relations for four different estimates of the specific stellar angular momentum are shown in Fig.\,\ref{fig:analysis/reliability/Fall relation HST comparison}. On the top left, the specific stellar angular momentum uses the bulge-removed \HST{} images in combination with the mass model rotation curve (i.e. similarly to Fig.\,\ref{fig:analysis/momentum first plot}). On the top right, it uses the bulge-removed \HST{} images combined with a flat rotation curve model. On the bottom-left, it uses the best-fit exponential disk mass distribution (derived with \Galfit{}) in combination with the mass model rotation curve (the angular momentum integrated numerically, see Eq.\,\ref{eq:derivation/specific_angular_momentum_this_analysis}). And, on the bottom-right, is uses an exponential disk mass distribution combined with a flat model rotation curve (the angular momentum is evaluated analytically, see Appendix\,\ref{appendix:angular momentum}). The symbols are the same as in Fig.\,\ref{fig:sample/selection/selection}. 
	
	We start by considering the effect of the rotation curve. Comparing the plots on the first row of Fig.\,\ref{fig:analysis/reliability/Fall relation HST comparison}, we see large variations of the specific stellar angular momentum. Overall, the difference between the estimate that uses the mass model and the one that uses the flat model rotation curve is roughly $0.35 \pm \SI{0.35}{\dex}$ for the entire kinematics sample. This becomes $0.1 \pm \SI{0.15}{\dex}$ when considering the selected sample only and it is amplified to $0.45 \pm \SI{0.4}{\dex}$ when considering the galaxies removed by the selection. Galaxies flagged as having peculiar kinematics are also greatly impacted with a difference of around $0.3 \pm \SI{0.25}{\dex}$, in between the samples of galaxies kept and removed by the selection. In particular, we see that flagged galaxies tend to be aligned along a lower limiting line (as indicated by the black lines in Fig.\,\ref{fig:analysis/reliability/Fall relation HST comparison}) when using the mass model rotation curve, whereas they are spread throughout and below the relation when using a flat model rotation curve. As discussed in the previous section, the reason for that effect is that most of these galaxies lack a visible velocity gradient (e.g. \textsc{70-CGr79} in Fig.\,\ref{fig:sample/selection/examples dynamical models}). Thus, it is unlikely that their kinematics is properly fitted, meaning that the contributions of their stellar disk and bulge components are overestimated and that no DM halo could be fitted, hence producing a spurious correlation between the galaxies' stellar mass and specific angular momentum\footnote{The reason for this correlation is that the stellar mass is linked to the galaxies' stellar mass distribution. In the case of a DM halo-free galaxy, the gravitational potential is dominated by that of the stars, so that, under the assumption of equilibrium, the kinematics is also directly governed by the stellar mass distribution.}. Because of varying bulge contributions to the rotation curve and non-axisymmetric features in the \HST{} images (e.g. clumps), this effect is not systematic for every flagged galaxy (e.g. 17\% of them have a stellar angular momentum difference below 15\%). Then, we investigate the effect of using \HST{} images instead of an exponential disk model. This is shown in Fig.\,\ref{fig:analysis/reliability/Fall relation HST comparison} by vertically comparing the plots in a same column (i.e. left-hand column for the mass model rotation curve and right-hand column for the flat rotation curve). We find that the use of \HST{} images only has a small impact on the position of the galaxies in the Fall relation. Selected galaxies are barely affected with a difference (angular momentum from HST images minus that from the exponential disk model) of $0.0 \pm \SI{0.1}{\dex}$, whereas unselected galaxies tend have slightly larger values with an average difference of roughly $-0.15 \pm \SI{0.25}{\dex}$. This can be explained by the fact that most of the unselected galaxies are removed by the surface selection criterion (blue upward pointing triangles and grey dots) which tends to mostly remove small galaxies (grey dots only). Since their best-fit disk's model is not as precisely constrained as larger galaxies, it is likely that it overestimates the stellar mass in the disk, therefore producing higher specific angular momentum values.
	
	Hence, these comparisons suggest that the choice of the rotation curve is what mainly drives the values of the specific stellar angular momentum. In particular, one must be careful when using the mass model rotation curve not to be biased by galaxies with large stellar mass fraction uncertainties (i.e. galaxies that seemingly appear baryon-dominated, though they are not intrinsically so) given that these objects produce spurious correlations at the bottom of the Fall relation (i.e. for low $j_\star$ values at fixed stellar mass). Because the same models and selection function were applied to the entire kinematics sample independently of the galaxies' environment, we do not expect this effect to be directly correlated to the density estimates. Thus, this should not affect our conclusions.
	
	\section{Results}
	\label{sec:results}
	
	We present in this section the results of the analysis of the impact of the environment. To begin with, we probe in Sect.\,\ref{sec:Analysis/environment/richness} the environment using the groups' richness. In Sect.\,\ref{sec:Analysis/environment/other estimators}, we consider two different environment tracers that probe the position of the galaxies in phase-space and the over-density around them. In Sect.\,\ref{sec:Analysis/environment/dv_sigmaV}, we discuss the link between the angular momentum of the galaxies and their normalised systemic velocity with respect to their host group. And, in Sect.\,\ref{sec:Analysis/environment/dispersion}, we discuss the link between the angular momentum of the galaxies and the velocity dispersion of their ionised gas component.
	
	\subsection{Environment traced by groups' richness}
	\label{sec:Analysis/environment/richness}
	
	\subsubsection{Comparison between galaxies in the field and in rich groups}	
	\label{sec:Analysis/environment/richness/results}
	
	\begin{figure}[htp]
		\centering
		\includegraphics[scale=0.9]{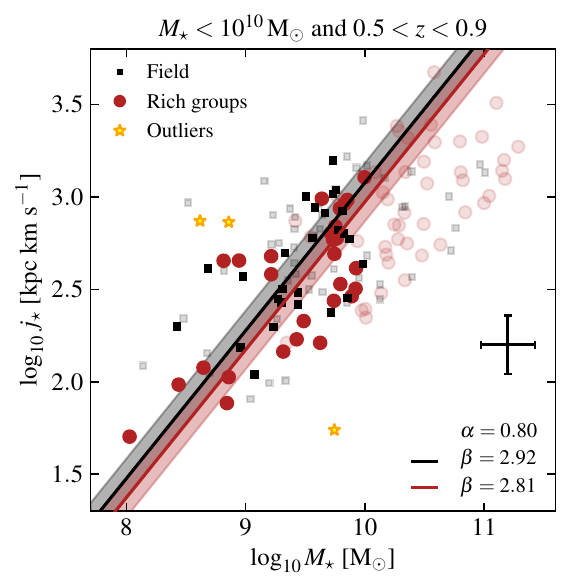}
		\caption{Fall relation for galaxies in the field (black squares) and those in rich groups with more than ten members (red circles). A stellar mass cut $M_\star < \SI{e10}{\Msun}$ and a redshift cut $0.5 < z < 0.9$ were applied to each sub-sample. Outliers detected during the fitting process are identified with yellow stars and the typical uncertainty is shown with a black error-bar. The best-fit lines are shown for the field (black) and groups' (red) sub-samples, along with their 95\% confidence intervals (coloured areas) determined with jackknife resampling. Galaxies removed by the mass and redshift cuts are identified with semi-transparent symbols.}
		\label{fig:analysis/environment/Fall relation field vs structures}
	\end{figure}

	First, we consider the impact of the environment using the simplest density tracer that we computed: the richness of the groups. This parameter was already used in \citet{mercier_scaling_2022} to study the impact of the environment on the size-mass, MS, and Tully-Fisher relations. We consider the two following sub-samples: 
\begin{enumerate*}[label=(\roman*)]
	\item field galaxies, that is galaxies that were not associated to any group by the FoF algorithm or that were associated to a group with less than three members, and
	\item galaxies located in rich groups that have at least ten galaxy members ($N>10$).
\end{enumerate*}
	Their Fall relations are illustrated in Fig.\,\ref{fig:analysis/environment/Fall relation field vs structures}. Field galaxies are shown with black points, those in groups with red circles, and outliers identified during the fitting process with yellow stars. The black line represents the best-fit linear relation for the field galaxies sample and the red line that for the galaxies in groups (both were obtained using the same slope derived on the entire sample). We also show the 95\% confidence intervals for both fits, estimated using jackknife resampling. In this figure, we apply a stellar mass cut $M_\star < \SI{e10}{\Msun}$ and a redshift cut $0.5 < z < 0.9$ in order not to be biased by massive galaxies that are almost entirely found in the richest groups (i.e. we do not have a statistically significant sample of field galaxies at $M_\star \gtrsim \SI{e10}{\Msun}$, for more details see the discussion in \citealp{mercier_scaling_2022}), as well as by a potential redshift evolution of the Fall relation\footnote{Galaxies in the groups are nearly all found at $z \approx 0.7$, whereas field galaxies are spread throughout the entire redshift range. The interested reader can find the resdhift and stellar mass distributions in Figs.\,1 and 2 of \citet{mercier_scaling_2022}.}. As an indication, we also show the galaxies removed by the two cuts with semi-transparent symbols. To gain deeper insights into the impact of the environment on the Fall relation, we also fitted it without any cut, with the stellar mass cut only, and with the redshift cut only. We also did the same but using the flat model rotation curve instead. The results of all these fits in terms of slope, best-fit zero-point, and confidence intervals are summarised in Table\,\ref{tab:fit_Fall_relation_structure_richness}.
	
	\begin{table}
		\centering
		\caption{Best-fit values for the Fall relation using the groups' richness as density estimate.}
		\resizebox{0.48\textwidth}{!}{%
		\begin{tabular}{c|llcccc}
		\hline\\[-9pt]
		& Cut & Sub-sample & Nb. galaxies & $\alpha$ & $\beta$ & 95\% CI\\ 
		& (1) & (2)        & (3)          & (4)      & (5)     & (6)\\
		\hline
		\hline
		\multirow{8}{*}{\rotatebox[origin=c]{90}{Mass model}} & \multirow{2}{*}{None} & Field & 85 & \multirow{2}{*}{0.50} & 2.80 & 2.73, 2.86\\
		& & Large & 75 & & 2.68 & 2.62, 2.74\\
		\cline{2-7}\\[-9pt]
		& \multirow{2}{*}{Mass} & Field & 67 & \multirow{2}{*}{0.83} & 2.96 & 2.89, 3.04\\
		& & Large & 36 & & 2.84 & 2.73, 2.95\\
		\cline{2-7}\\[-9pt]            
		& \multirow{2}{*}{Redshift} & Field & 33 & \multirow{2}{*}{0.52} & 2.76 & 2.67, 2.86\\
		& & Large & 66 & & 2.68 & 2.61, 2.75\\
		\cline{2-7}\\[-9pt]
		& \multirow{2}{*}{\bf{Both}} & \bf{Field} & \bf{27} & \multirow{2}{*}{\bf{0.80}} & \bf{2.92} & \bf{2.81, 3.02}\\
		& & \bf{Large} & \bf{30} & & \bf{2.81} & \bf{2.71, 2.92}\\
		\hline
		\hline
		\multirow{8}{*}{\rotatebox[origin=c]{90}{Flat model}} & \multirow{2}{*}{None} & Field & 85 & \multirow{2}{*}{0.55} & 2.73 & 2.65, 2.80\\
		& & Large & 75 & & 2.55 & 2.48, 2.62\\
		\cline{2-7}\\[-9pt]
		& \multirow{2}{*}{Mass} & Field & 67 & \multirow{2}{*}{0.94} & 2.92 & 2.83, 2.99\\
		& & Large & 36 & & 2.74 & 2.62, 2.87\\
		\cline{2-7}\\[-9pt]            
		& \multirow{2}{*}{Redshift} & Field & 33 & \multirow{2}{*}{0.57} & 2.66 & 2.54, 2.78\\
		& & Large & 66 & & 2.52 & 2.44, 2.60\\
		\cline{2-7}\\[-9pt]
		& \multirow{2}{*}{\bf{Both}} & \bf{Field} & \bf{27} & \multirow{2}{*}{\bf{0.91}} & \bf{2.86} & \bf{2.74, 2.99}\\
		& & \bf{Large} & \bf{30} & & \bf{2.74} & \bf{2.60, 2.88}\\
		\hline
		\end{tabular}}
		\label{tab:fit_Fall_relation_structure_richness}\vspace{5pt}
			{\small\raggedright {\bf Notes:} (1) Selection cut applied, (2) sub-sample (i.e. galaxies in the field or in rich groups), (3) number of galaxies in the sub-sample, (4) best-fit slope, (5) best-fit zero-point, and (6) jackknife resampling-derived 95\% confidence intervals for the best-fit zero-point. The same slope was used for both sub-samples. It was derived by fitting first with a free slope the entire sample with the considered selection cuts applied. The stellar mass cut selects galaxies according to $M_\star < \SI{e10}{\Msun}$ and the redshift cut according to $0.5 < z < 0.9$.\par}
	\end{table}
	
	Without any cut, we obtain a $3.5\sigma$-significant difference (with $\sigma$ the jackknife-derived standard-deviation) in zero-points of \SI{0.12}{\dex} between galaxies in the field and those in rich groups, consistent with a depletion of specific stellar angular momentum at fixed stellar mass for the galaxies in the groups with respect to the field. When applying both the stellar mass and the redshift cuts, the zero-points of the two sub-samples increase each by roughly \SI{0.13}{\dex} (i.e. the difference remains the same), but their uncertainties increase as well (changing from \SI{0.03}{\dex} to \SI{0.05}{\dex}), so that the difference in zero-points reduces to $2\sigma$-significant only. Thus, it is not straightforward to conclude whether the observed difference is statistically induced or environmentally-driven. Especially when applying both the mass and redshift cuts, we cannot completely rule out the null hypothesis that both sub-samples share the same zero-point. Besides, independently of whether we apply the cuts or not, removing galaxies flagged as having peculiar kinematics (identified with red contours in Figs.\,\ref{fig:analysis/momentum first plot} and \ref{fig:analysis/reliability/Fall relation HST comparison}) has a negligible impact on the results (up to \SI{0.02}{\dex}). We also find the same trend, that is a depletion of specific stellar angular momentum for the galaxies in the groups with respect to the field, when using the flat model rotation curve (see Fig.\,\ref{fig:appendix/plots/Fall relation env flat}). Compared to the mass models, the zero-point changes by \SI{-0.06}{\dex} for either sub-sample. Without any cuts, we get a difference in zero-points of \SI{0.22}{\dex} ($4.5\sigma$-significant), which reduces to \SI{0.12}{\dex} (around $2\sigma$-significant) when applying both stellar mass and redshift cuts.
	
	We note that this observed depletion of angular momentum for galaxies located in rich groups with respect to the field is compatible with previous results from both \citet{pelliccia_searching_2019} and \citet{Perez2021}. In particular, looking at Fig.\,8 and Table\,5 of \citet{Perez2021}, we determine that they found a similar depletion of angular momentum of around $0.1 - \SI{0.15}{\dex}$ between their sample of galaxies located in clusters and the various field samples they used for comparison. Similarly, \citet{pelliccia_searching_2019} also found a \SI{0.1}{dex} depletion in groups/clusters compared to the field (see their Fig.\,8 and Sect.\,4.4).
	
	\subsubsection{Effect of stellar mass and redshift}
	\label{sec:Analysis/environment/richness/cuts}

	It might be interesting to comment on the effect of stellar mass and redshift cuts. First, the mass cut increases the slope from $\alpha = 0.5$ when using the mass models ($\alpha = 0.55$ for the flat model) to $\alpha = 0.83$ after the cut is applied ($\alpha = 0.94$). It also increases the zero-points of the field sub-sample by \SI{0.2}{\dex} (\SI{0.25}{\dex} for the flat model) and of the rich groups sub-sample by \SI{0.14}{\dex} (\SI{0.21}{\dex}). The main impact of this change of slope is to bring the most massive galaxies in the sample ($M_\star \gtrsim \SI{e10}{\Msun}$) below the best-fit lines obtained when applying either the redshift cut only or no cut at all. Thus, massive galaxies appear depleted in angular momentum with respect to what would be expected by extrapolating the Fall relation found at lower mass, as is visible in Fig.\,\ref{fig:analysis/environment/Fall relation field vs structures}. In particular, the more massive the galaxies, the lower is their average angular momentum (up to \SI{-0.5}{dex} with respect to the best-fit line, see Table.\,\ref{tab:depletion_momentum_richness} for a complete list of values). However, the lack of a statistically significant sample of field galaxies in this mass range prevents us from drawing conclusions as to whether this effect is mass- or environmentally-driven (or both). Furthermore, this effect might also be statistically induced by the fact that, when applying the mass cut, we are giving more weight to low-mass galaxies with larger uncertainties and with a coarser sampling. 
	
	On the contrary, the redshift cut has nearly no effect on the slope (changing it by $\pm \SI{0.03}{\dex}$ at most) and on the zero-point of the sub-sample of galaxies in rich groups. However, it does reduce that of the field galaxies sub-sample (\SI{-0.04}{\dex} with the redshift cut and \SI{-0.07}{\dex} with both mass and redshift cuts). Galaxies located in rich groups are not affected by the redshift cut mostly because more than 90\% are located in the range $0.5 < z < 0.9$. The visible effect of the redshift cut on the field sub-sample might be an indication of a redshift evolution of the Fall relation's zero-point. This is strengthened by the fact that around 65\% of galaxies at $z > 0.9$ are found above the best-fit Fall relation whereas galaxies at $z < 0.5$ are located evenly throughout and along the relation. However, given the low statistics at $z < 0.5$ and $z > 0.9$ (about 20 galaxies each), it is difficult to assess whether this effect is statistically-induced or due to an evolution of the zero-point of the Fall relation with redshift. In particular, massive galaxies that are depleted in angular momentum (i.e. located below the best-fit line) are mostly found at $z < 0.9$, meaning that they will mainly lower the zero-point for samples that are at $z < 0.9$ and give the impression that galaxies at $z > 0.9$ have higher angular momentum values.

	\subsection{Link with global and local density estimates}
	\label{sec:Analysis/environment/other estimators}
	
	\begin{figure}
		\centering
		\includegraphics[scale=0.9]{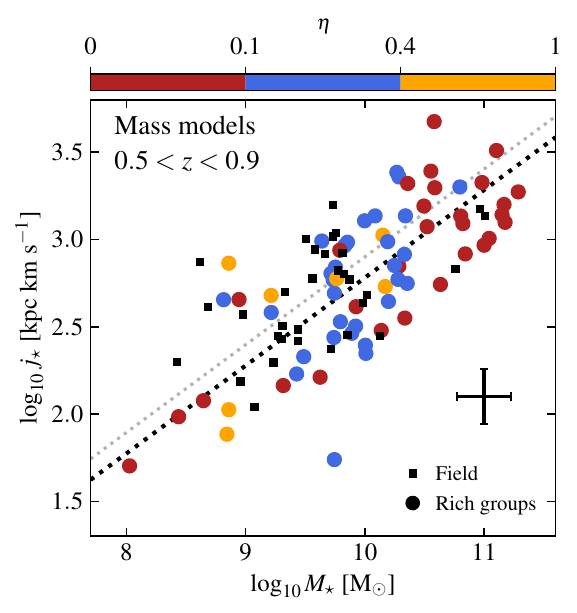}
		\caption{Fall relation colour coded by $\eta$ (see Sect.\,\ref{sec:sample/environment}). Galaxies in rich groups with more than ten members are shown as circles and field galaxies as black squares (both have the redshift cut $0.5 < z < 0.9$ applied). The mass model rotation curve is used (see Fig.\,\ref{fig:appendix/plots/Fall relation eta flat} for the flat model rotation curve instead). As an indication, the best-fit linear relations (with redshift cut applied, see also Table\,\ref{tab:fit_Fall_relation_structure_richness}) for the field and rich groups sub-samples are shown as grey and black plain lines, respectively. The typical uncertainty is shown as the black error-bar on the bottom right corner.}
		\label{fig:analysis/environment/Fall relation eta}
	\end{figure}
	
%	
%	
%	\begin{figure*}
%		\centering
%		\includegraphics[scale=0.72]{momentum_mass_in_R22_normR22_vs_eta}
%		\caption{Fall relation colour coded according to $\eta$ (left panel), the galaxies' line-of-sight systemic velocity $\Delta v$ normalised by the velocity dispersion of their host group $\sigma_V$ (middle panel), and the galaxies' distance to the centre of their host group normalised by the virial radius $R_{200}$ of the group (right panel). Only galaxies located in rich groups with more than ten members are shown as circles (black points correspond to field galaxies). All plots use the mass model rotation curve with the redshift cut ($0.5 < z < 0.9$) applied (see Fig.\,\ref{fig:appendix/plots/Fall relation eta flat} for the flat model rotation curve instead). As an indication, the best-fit linear relations for the field and rich groups sub-samples with the stellar mass cut ($M_\star < \SI{e10}{\Msun}$) applied are shown as grey and black plain lines, respectively, and without the mass cut applied as grey and black dotted lines, respectively. The typical uncertainty is shown as the black errorbar on the bottom right corner of the right panel.}
%		\label{fig:analysis/environment/Fall relation eta}
%	\end{figure*}
	
	We refine the previous analysis by using more precise density estimates. Indeed, the groups' richness is a simple proxy to probe the effect of the environment and, as such, it potentially mixes galaxies located in regions of various densities that might have been affected by their environment on different time-scales. To do so, we use the global density estimate $\eta$ that encompasses in a single parameter the position and dynamics of the galaxies in their host group (see Sect.\,\ref{sec:sample/environment}). Following \citet{Noble2013}, we separate galaxies associated to groups between those that have been accreted early ($\eta < 0.1 $), "backsplash" galaxies that have already passed the pericentre of their orbit once \citep[$0.1 < \eta < 0.4$, e.g.][]{Balogh2000, Gill2005}, and recently accreted galaxies infalling into the groups ($0.4 < \eta < 2$)\footnote{We do not have galaxies with $\eta > 2$ because all these objects have already been associated to the field sub-sample by the FoF algorithm.}. In our case, we lack statistics across the entire stellar mass range in \MAGIC{} to properly sample
\begin{enumerate*}[label=(\roman*)]
	\item early accreted galaxies when applying both mass and redshift cuts (eight galaxies remaining) and
	\item infalling galaxies ($\eta > 0.4$) when applying either the mass cut only (13 galaxies remaining) or both mass and redshift cuts (seven galaxies remaining). 
\end{enumerate*}
	Second, as can be seen in Fig.\,\ref{fig:analysis/environment/Fall relation eta}, different $\eta$ classes populate different mass ranges, meaning that we would be biased by the different mass distributions when fitting each class separately. Thus, instead of fitting these classes, we compare their distributions in the $j_\star - M_\star$ plane with the field sample and with respect to the best-fit lines (with redshift cut applied, see Table\,\ref{tab:fit_Fall_relation_structure_richness}) obtained in Sect.\,\ref{sec:Analysis/environment/richness}. The Fall relation for these three $\eta$ classes is shown in Fig.\,\ref{fig:analysis/environment/Fall relation eta} with early accreted galaxies represented as red circles, backsplash galaxies as blue circles, infalling galaxies as orange circles, and field galaxies as black points. We note that $\eta$ can also be computed for smaller groups, but including the galaxies from these groups does not affect the conclusions. Thus, to remain consistent with previous sections we only focus on rich groups in what follows.
	
	More than 90\% of the backsplash galaxies ($0.1 < \eta < 0.4$) are in the stellar mass range $10^9 < M_\star / \unit{\Msun} < 10^{10.5}$ and around 75\% of the early accreted galaxies ($\eta < 0.1$) have $M_\star > \SI{e10}{\Msun}$ (more than 50\% above $10^{10.5}\,\unit{\Msun}$). Thus, there is a strong dichotomy between these two classes which seems to be mostly driven by stellar mass, with a transition at around \SI{e10}{\Msun}. Overall, 65\% of the early accreted galaxies are located below the best-fit line (with redshift cut applied, see Table\,\ref{tab:fit_Fall_relation_structure_richness}), whereas infalling galaxies are spread evenly throughout. This number rises to 75\% when compared to the best-fit line obtained when applying both mass and redshift cuts. 
	
	Hence, early accreted galaxies are located on average below the best-fit Fall relation. If we separate the sub-sample in various mass bins, we see that low-mass galaxies ($M_\star < 10^{10.5}\,\unit{\Msun}$) are mostly associated with backsplash galaxies ($\eta \approx 0.2$), whereas massive galaxies are nearly all identified as early accreted with $\eta < 0.1$ (see Table\,\ref{tab:eta} for the complete list of values). In other words, low $\eta$ values for galaxies located at the bottom of the Fall relation are mostly driven by massive galaxies. These results also hold when using the flat model rotation curve (see Fig.\,\ref{fig:appendix/plots/Fall relation eta flat}).
	
	\begin{table}
		\centering
		\caption{Global and local density in various stellar mass bins for the sub-sample of galaxies located in rich groups.}
		\resizebox{0.48\textwidth}{!}{%
		\begin{tabular}{cccc}
		\hline\\[-9pt]
		Nb. & Stellar mass bin & $\eta$ & Local over-\\ 
		galaxies & [$\log_{10} \unit{\Msun}$] & & density $\delta$\\
		(1) & (2) & (3) & (4)\\
		\hline
		\hline\\[-2ex]
		13 & 8.5, 9.5 & $0.19^{0.71}_{0.09}$ & $0.81^{1.00}_{0.40}$ \\[1ex]
		22 & 9, 10 & $0.20^{0.34}_{0.10}$ & $0.82^{1.10}_{0.48}$ \\[1ex]
		37 & 9.5, 10.5 & $0.17_{0.09}^{0.31}$ & $0.86^{1.11}_{0.54}$ \\[1ex]
		31 & 10, 11 & $0.10_{0.04}^{0.24}$ & $0.98^{1.34}_{0.64}$ \\[1ex]
		18 & 10.5, 11.5 & $0.04_{0.04}^{0.06}$ & $1.19^{1.48}_{0.83}$ \\[0.5ex]
		\hline
		\end{tabular}}
		\label{tab:eta}\vspace{5pt}
			{\small\raggedright {\bf Notes:} (1) Number of galaxies located in rich groups in the given stellar mass range, (2) minimum and maximum bounds of the stellar mass bin, (3) median, 16th, and 84th percentiles of $\eta$, and (4) median, 16th (subscript), and 84th (superscript) percentiles of the VMC-based over-density $\delta$. For definitions of $\eta$ and $\delta$, see Sect.\,\ref{sec:sample/environment}.\par}
	\end{table}
	
	We also investigate whether this dichotomy is observed with a direct tracer of the over-density around the galaxies. To do so, we measured the median value of the VMC-based over-density $\delta$ (see Sect.\,\ref{sec:sample/environment}) for low- and high-mass galaxies (see Table.\,\ref{tab:eta} for the complete list of values and Figs.\,\ref{fig:appendix/plots/Fall relation Voronoi Brian} and \ref{fig:appendix/plots/Fall relation Voronoi Brian flat} for the Fall relation colour-coded with $\delta$). We do find an increase from $\delta \approx 5$  for $M_\star < \SI{e10}{\Msun}$ to $\delta \approx 9$ above this mass threshold. However, when splitting the sample between galaxies located below and above the best-fit Fall relation (with redshift cut applied, see Table\,\ref{tab:fit_Fall_relation_structure_richness}), we do not find any significant difference in the distributions of $\delta$. Moreover, this result is independent of the choice of rotation curve.
	
	Hence, there is no evidence that the depletion of angular momentum for galaxies in the groups with respect to the field is associated with the over-density around the galaxies. Furthermore, even though galaxies located below the Fall relation tend to have a larger fraction of early accreted galaxies, the effect seems to be dominated by their stellar mass rather than the environment.
	
	\subsection{Motion-induced loss of angular momentum ?}
	\label{sec:Analysis/environment/dv_sigmaV}
	
	A potential caveat of using $\eta$ in Sect.\,\ref{sec:Analysis/environment/other estimators} is that it combines in a single parameter the position with the velocity of the galaxies in their host group but, depending on the underlying physical mechanism, each can contain different information (e.g. ram-pressure stripping primarily scales with velocity). Thus, in this section, we split $\eta$ into a velocity component $| \Delta v | / \sigma_V$, where $\Delta v$ is the systemic velocity of a galaxy with respect to the dynamical centre of its host group and $\sigma_V$ is the velocity dispersion of the group (see Sect.\,\ref{sec:sample/environment}), and a distance component $R/R_{200}$, where $R_{200}$ is the virial radius of the group. 
	
	To begin with, we do not find any obvious correlation between the position of the galaxies along, above, or below the Fall relation and their normalised distance $R/R_{200}$. However, we do find that the most massive galaxies ($M_\star \gtrsim 10^{10.5}\,\unit{\Msun}$) in the groups are primarily found in the inner parts of the groups (on average $R/R_{200} \approx 0.15$ above this mass threshold and $R/R_{200} \approx 0.3$ below). We note that this is in line with previous results found for galaxy groups in the local Universe \citep[e.g.][]{Roberts2015}.
	
\begin{figure}
		\centering
		\includegraphics[scale=0.9]{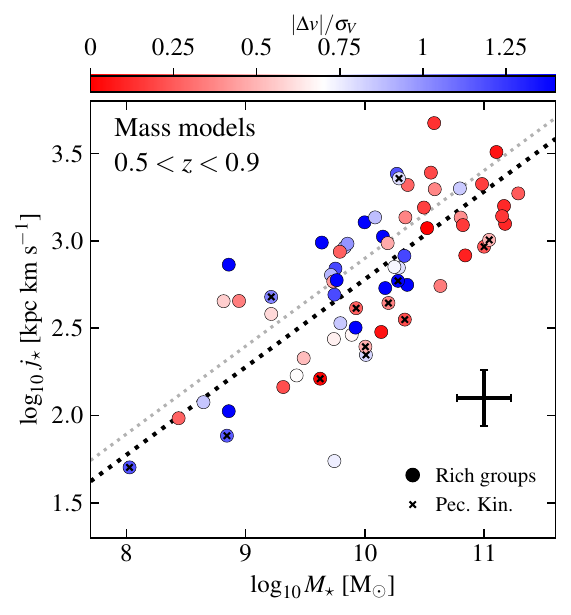}
		\caption{Fall relation for galaxies in rich groups (with more than ten members) colour coded by the galaxies' line-of-sight systemic velocity $|\Delta v|$ normalised by the velocity dispersion of their host group $\sigma_V$. Galaxies with peculiar kinematics identified in Sect.\,\ref{sec:sample/selection} are shown with a black cross. The mass model rotation curve is used (see Fig.\,\ref{fig:appendix/plots/Fall relation dV_sigma flat} for the flat model rotation curve instead). As an indication, the best-fit linear relations (with redshift cut applied, see also Table\,\ref{tab:fit_Fall_relation_structure_richness}) for the field and rich groups sub-samples are shown as grey and black plain lines, respectively. The typical uncertainty is shown as the black error-bar on the bottom right corner.}
		\label{fig:analysis/environment/Fall relation dV_sigma}
	\end{figure}

	On the other hand, we do find a significant correlation between
\begin{enumerate*}[label=(\roman*)]
	\item $| \Delta v | / \sigma_V$,
	\item the stellar mass of the galaxies, and
	\item their position orthogonal to the Fall relation (i.e. above or below it).
\end{enumerate*}	
	The Fall relation colour-coded by $| \Delta v | / \sigma_V$ is shown in Fig.\,\ref{fig:analysis/environment/Fall relation dV_sigma} for the mass model rotation curve and in Fig.\,\ref{fig:appendix/plots/Fall relation dV_sigma flat} for the flat model rotation curve (both for galaxies in rich groups only). First, we find that massive galaxies ($M_\star \gtrsim 10^{10.5}\,\unit{\Msun}$) tend to have the lowest $| \Delta v | / \sigma_V$ values with, on average, $| \Delta v | / \sigma_V \approx 0.17$  and 0.8 for galaxies above and below this mass threshold, respectively. This dichotomy in $| \Delta v | / \sigma_V$ strongly correlates with the dichotomy seen for $\eta$. Thus, this means that the massive early accreted galaxies located at the high-mass end are so mostly because of their low systemic velocity with respect to the velocity dispersion (or alternatively dynamical mass) of their host group.
	
	 Interestingly and as clearly visible in Fig.\,\ref{fig:analysis/environment/Fall relation dV_sigma}, we also find a strong dichotomy in $| \Delta v | / \sigma_V$ at lower stellar masses ($M_\star \lesssim 10^{10.5}\,\unit{\Msun}$), but this time orthogonal to the Fall relation. Overall, galaxies above the best-fit line (when applying the redshift cut only) have a median $|\Delta v| / \sigma_V$ equal to 0.9, whereas it is only 0.7 below. Furthermore, we also find that around 70\% of the galaxies with $| \Delta v | / \sigma_V < 0.7$ are located below this line. When further splitting $| \Delta v | / \sigma_V$ into the systemic velocity of the galaxies $|\Delta v|$ and their host group's dispersion $\sigma_V$, we find that this dichotomy is mostly driven by the systemic velocity of the galaxies. In other words, galaxies located below the Fall relation appear to have lower systemic velocities than those located above it. This includes the majority of the low-mass early accreted galaxies ($\eta < 0.1$) for which we expect, by definition, to have relatively low $|\Delta v| / \sigma_V$ values, as well as some of the backsplash galaxies ($0.1 < \eta < 0.4$). This result suggests that the mechanism responsible for the loss of specific stellar angular momentum for the galaxies located in the groups must be tightly linked to the velocity at which galaxies move through the intra-group medium.
	
	\subsection{An anti-correlation between gas dispersion and angular momentum}
	\label{sec:Analysis/environment/dispersion}
	
\begin{figure}
		\centering
		\includegraphics[scale=0.9]{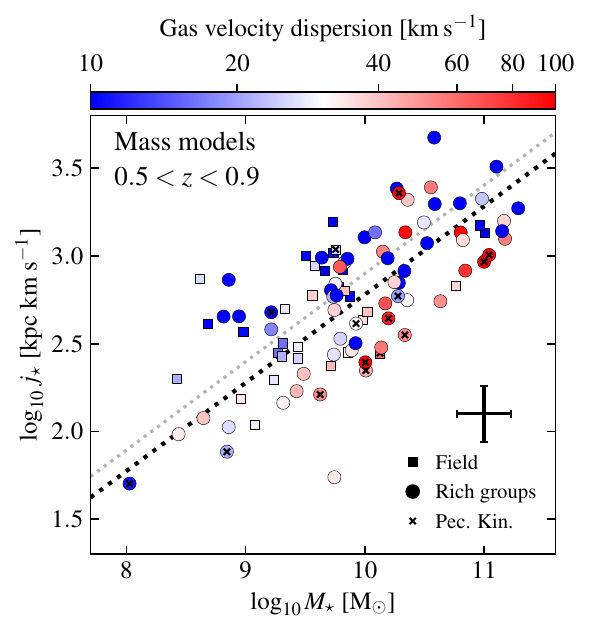}
		\caption{Fall relation colour coded according to the median velocity dispersion of the ionised gas component in the galaxies. Galaxies in rich groups with more than ten members are shown as circles and field galaxies as smaller squared (both have the redshift cut $0.5 < z < 0.9$ applied). Galaxies with peculiar kinematics identified in Sect.\,\ref{sec:sample/selection} are shown with a black cross. The mass model rotation curve is used (see Fig.\,\ref{fig:appendix/plots/Fall relation dV_sigma flat} for the flat model rotation curve instead). As an indication, the best-fit linear relations (with redshift cut applied, see also Table\,\ref{tab:fit_Fall_relation_structure_richness}) for the field and rich groups sub-samples are shown as grey and black plain lines, respectively. The typical uncertainty is shown as the black error-bar on the bottom right corner.}
		\label{fig:analysis/environment/Fall relation gas dispersion}
	\end{figure}

	We complete this analysis by investigating whether there is a link between the depletion of specific stellar angular momentum for galaxies in rich groups with respect to the field (see Sect.\,\ref{sec:Analysis/environment/richness}) and the velocity dispersion of the gas. Finding such a link would be interesting because it could pinpoint at mechanisms that are able to either increase or decrease the dispersion of the gas as the galaxies move through the groups (e.g. mergers or thermal evaporation). The Fall relation colour-coded according to the median value of the galaxies' ionised gas velocity dispersion\footnote{Measured from the beam-smearing- and LSF-corrected velocity dispersion map extracted from the \MUSE{} cubes.} is shown in Fig.\,\ref{fig:analysis/environment/Fall relation gas dispersion} for the mass model rotation curve and in Fig.\,\ref{fig:appendix/plots/Fall relation gas dispersion flat} for the flat model rotation curve.
	
	Similarly to $|\Delta v | / \sigma_V$, we also find a strong dichotomy in terms of the ionised gas velocity dispersion orthogonal to the Fall relation. Galaxies located below the best-fit line (when applying the redshift cut only) have an average gas velocity dispersion of around \SI{40}{\kilo\meter\per\second}\footnote{We note that the typical LSF dispersion is around 20-\SI{30}{\kilo\meter\per\second}}, whereas those located above it rather have values around \SI{10}{\kilo\meter\per\second}. Furthermore, 75\% of the galaxies above the best-fit line have an ionised gas velocity dispersion below \SI{25}{\kilo\meter\per\second} (hence the low median value), whereas 75\% of the galaxies below the best-fit line have values above \SI{25}{\kilo\meter\per\second}, with a peak at around \SI{50}{\kilo\meter\per\second}. It is also interesting to note that, contrary to what was found in Sect.\,\ref{sec:Analysis/environment/dv_sigmaV}, this dichotomy is not correlated with the stellar mass of the galaxies. Hence, massive early accreted galaxies discussed in the previous sections do not appear to have either higher or lower gas velocity dispersion values than lower mass galaxies.
	
	Thus, galaxies located above the Fall relation appear to be strongly rotationally-supported with very low ionised gas velocity dispersion, whereas galaxies located below the Fall relation have significantly higher values. This result is not entirely new and a similar correlation was already observed in previous analyses that used the dynamical state of the galaxies $V/\sigma$ rather than the velocity dispersion of the gas alone (e.g. \citealp{contini_deep_2016}, but in particular \citealp{bouche_muse_2021} where this correlation was highlighted). However, these studies focussed mostly on field galaxies, therefore pointing to an underlying mechanism that is not environmentally-induced. This is further supported by the fact that a similar dichotomy can be found for field galaxies in our sample. More than 70\% of the field galaxies located above the Fall relation have ionised gas velocity dispersion values below \SI{25}{\kilo\meter\per\second}, whereas roughly 70\% of those below have a gas velocity dispersion above \SI{25}{\kilo\meter\per\second}. Still, we do see an enhancement of the ionised gas velocity dispersion for galaxies in the groups. In particular, we find that more than 30\% of the galaxies located in rich groups have a ionised gas velocity dispersion above \SI{50}{\kilo\meter\per\second}, whereas only 5\% of the field galaxies show a dispersion this high.
	
	\subsection{Summary and interpretation of the results}
	\label{sec:Analysis/interpretation}
	
	Interpreting these results in terms of underlying driving physical mechanisms that are either linked to the mass of the galaxies or to the environment around them is difficult without relying on hydrodynamical simulations, or at least analytical models of galaxy evolution. Our aim with the second paper of this series is actually to investigate how different physical processes can affect galaxies and move them along or across the Fall relation. Still, given the visible impact of the environment found in the previous sections and the interesting new correlations with the galaxies systemic velocity and gas velocity dispersion, we can propose a few scenarios that might produce such effects.
	
	In the previous sections, we have found that galaxies located in rich groups tend to be depleted in specific stellar angular momentum with respect to the field, especially at low mass where we have a comparable field sample. This depletion does not appear correlated neither with the position of the galaxies in their host group, nor with the over-density around them. However, low-mass galaxies located below the Fall relation show lower values of $|\Delta v | / \sigma_V$ than galaxies located at the top of the relation (i.e. they move the slowest in their host group) and they also have significantly higher ionised gas velocity dispersion values. Even though field galaxies show a comparable dichotomy in terms of gas velocity dispersion, there is nevertheless evidence that the groups might enhance it.
	
	Physical mechanisms compatible with these results must therefore be able to reduce the angular momentum either by changing the morphology of the galaxies (i.e. transferring mass from high to low circular velocity regions) or by decreasing the amplitude of the circular velocity (or alternatively change the shape of the rotation curve). We did show in \citet{mercier_scaling_2022} that \MAGIC{} galaxies located in rich groups are on average more compact than galaxies in the field at the same redshift (by roughly \SI{0.03}{\dex}). However, early calculations about the impact of such a difference on the Fall relation have convinced us that the offset is certainly too small to explain the large difference observed in Sect.\,\ref{sec:Analysis/environment/richness}. Furthermore, we also showed in \citet{mercier_scaling_2022} that low-mass ($M_\star < \SI{e10}{\Msun}$) galaxies in groups and in the field have similar circular velocities, meaning that the impact of the environment is certainly due to variations in the shapes of the rotation curves. This effect will be investigated in much more details in the second paper of the series.
	
	Since our stellar angular momentum estimate uses the ionised gas velocity as a proxy for the stellar kinematics, it should be sensitive to hydrodynamical mechanisms. The clear correlation with $|\Delta v | / \sigma_V$ presented in Sect.\,\ref{sec:Analysis/environment/dv_sigmaV}, the enhanced velocity dispersion of the ionised gas seen for galaxies in the groups, and the lack of correlation with the strength of the over-density around the galaxies could be an indication that we are seeing the effect of such physical processes (e.g. stripping or evaporation). This is strengthened by the fact that hydrodynamical mechanisms scale with the density of the intra-group medium rather than the density of galaxies and that some mechanisms, such as ram-pressure stripping, also scale with the velocity of the galaxies. Assuming there is enough pressure in the intra-group medium, these physical processes could affect the kinematics of the gas in the galaxies and, therefore, their angular momentum. On the other hand, gravitational interactions, in particular mergers, might also be an acceptable explanation. Indeed, they would be more likely to happen for galaxies that move slowly in the groups, they would affect both the morphology (contraction of baryons) and the kinematics of the gas and the stars, in particular enhancing their velocity dispersion. Furthermore, mergers could also be compatible with the lack of correlation with the strength of the over-density around the galaxies discussed in Sect.\,\ref{sec:Analysis/environment/other estimators} as there is some observational evidence \citep[e.g.][]{Shen2021} that such physical processes can locally reduce the density after the merging event.

	\section{Conclusions}
	\label{sec:conclusions}
	
	We have studied the dependence of the specific stellar angular momentum with environment using the Fall relation for nearly 200 galaxies at $z \sim 0.7$ from the \MAGIC{} survey. These include galaxies in the field and found in groups of various densities. Their stellar angular momentum was estimated with a new method that combines robust rotation curves of the ionised gas obtained with \MUSE{}, and used as a proxy of the stellar kinematics, with high-resolution \HST{} images. In particular, this approach allows us to alleviate the assumption of axial symmetry for the stellar disk component.
	
	We have shown that there is a visible impact of the environment on the angular momentum of galaxies located in rich groups with respect to the field. This effect is compatible with a depletion of angular momentum in the groups with respect to the field of around \SI{-0.12}{\dex}, which is $2\sigma$ significant after accounting for biases induced by different mass and redshift distributions between sub-samples. This first result is compatible with previous studies performed on galaxy clusters rather than groups, such as \citet{pelliccia_searching_2019} or \citet{Perez2021}. Furthermore, we find that massive galaxies have much lower specific stellar angular momentum values than what would be expected from extrapolating at the high-mass end the Fall relation found at low mass. However, we cannot rule out the possibility that this effect is statically induced rather than mass- or environmentally-driven. In particular, we will need to combine \MAGIC{} with other \MUSE{}-GTO surveys, such as \MUSCATEL{}\footnote{ESO ID 1104.A-0026}, \MUSEWIDE{} \citep{Herenz2017, Urrutia2019}, \HDFS{} \citep{bacon_muse_2015}, or \HUDF{} \citep{bacon_muse_2017, Bacon2023}, to increase the statistics at the high-mass end for the field sample. 
	
	When probing in more details the environment around the galaxies, we find little evidence for a correlation between their stellar angular momentum and the over-density around them, meaning that the observed depletion of angular momentum does not seem primarily driven by the local density in the groups. However, we do find that galaxies located below the Fall relation tend to be slightly more dynamically bound (as traced by $\eta$, see Sect.\,\ref{sec:sample/environment}) to their host group than those located above. Nevertheless, the effect appears to be mostly mass-driven with nearly all massive galaxies ($M_\star \gtrsim 10^{10.5}\,\unit{\Msun}$) identified as early accreted galaxies ($\eta < 0.1$). 
	
	We also looked at whether the depletion of angular momentum in the groups with respect to the field is associated to the position of the galaxies in their host group or to their systemic velocity (normalised by their host group velocity dispersion) with respect to the group's redshift. Even though we do not find any evidence that the position of the galaxies is linked to their angular momentum, we find a strong dichotomy in terms of systemic velocity between galaxies located below and above the Fall relation. Those below tend to have a significantly lower systemic velocity than those located above, with $|\Delta v| / \sigma_V \approx 0.7$ and 0.9 for those below and above, respectively, and with 70\% of the galaxies with $|\Delta v| / \sigma_V < 0.7$ found on the bottom of the Fall relation. This result hints at physical mechanisms that scale with the velocity at which galaxies move through the intra-group medium.
	
	Finally, we also find evidence for a strong link between the position of the galaxies below and above the Fall relation and their ionised gas velocity dispersion. Galaxies located above the relation appear to be strongly rotationally supported, with a median velocity dispersion of around \SI{10}{\kilo\meter\per\second}, whereas galaxies located below the relation have a higher median velocity dispersion value of about \SI{40}{\kilo\meter\per\second}. Nevertheless, we find a similar correlation using field galaxies, thus highlighting the fact that this effect might not be solely environmentally-driven. Still, the lower values found in the field sub-sample point to a scenario where the groups enhance the velocity dispersion of the gas as the galaxies lose angular momentum.
	
	From these results, we infer that the most likely pathway through which galaxies in groups lose partly their stellar angular momentum is by a change of the shape of their rotation curve rather than by reducing their maximum circular velocity or by redistributing their stellar mass. Both hydrodynamical processes, in particular stripping, and gravitational interactions, in particular mergers, can be viable scenarios that could produce both the depletion of angular momentum in the groups with respect to the field and the aforementioned correlations with $|\Delta v| / \sigma_V$ and with the ionised gas velocity dispersion.

	\begin{acknowledgements}
		We thank the referee for providing useful and constructive comments on the submitted version of this paper. 
		We dedicate this article in memory of Hayley Finley. 
		We acknowledge David Carton for his investment in the build-up of the project.
		This work was supported by the Programme National Cosmology et Galaxies (PNCG) of CNRS/INSU with INP and IN2P3, co-funded by CEA and CNES. 
		This work has been carried out through the support of the ANR FOGHAR (ANR-13-BS05-0010-02), the OCEVU Labex (ANR-11-LABX-0060), and the A*MIDEX project (ANR-11-IDEX-0001-02), which are funded by the “Investissements d’avenir” French government program managed by the ANR.
		This work has been carried out thanks to the support of the Ministry of Science, Technology and Innovation of Colombia (MINCIENCIAS) PhD fellowship program No. 756-2016.
		This research has made use of \matplotlib{} \citep{matplotlib}, \scipy{} \citep{scipy}, \numpy{} \citep{numpy} and \astropy{} \citep{astropy:2013, astropy:2018}.
	\end{acknowledgements}
	
	%%%%%%%%%%%%%%%%%%%%%%%%%%
	%      BIBLIOGRAPHY      %
	%%%%%%%%%%%%%%%%%%%%%%%%%%
	
	\bibliographystyle{aa}
	\bibliography{citations}
	
	%%%%%%%%%%%%%%%%%%%%%%%%
	%      APPENDICES      %
	%%%%%%%%%%%%%%%%%%%%%%%%
	
	\begin{appendix}

	\section{Additional figures and tables}
	
	\begin{figure}[htp]
		\centering
		\includegraphics[scale=0.9]{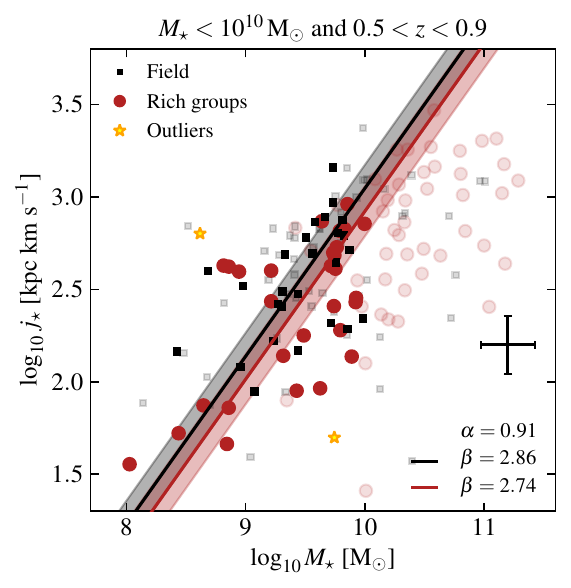}
		\caption{Fall relation between galaxies in the field (black squares) and those in rich groups with more than ten members (red circles) using the flat model rotation curve. See Fig.\,\ref{fig:analysis/environment/Fall relation field vs structures} for details regarding the legend.}
		\label{fig:appendix/plots/Fall relation env flat}
	\end{figure}
	
	\begin{figure}
		\centering
		\includegraphics[scale=0.9]{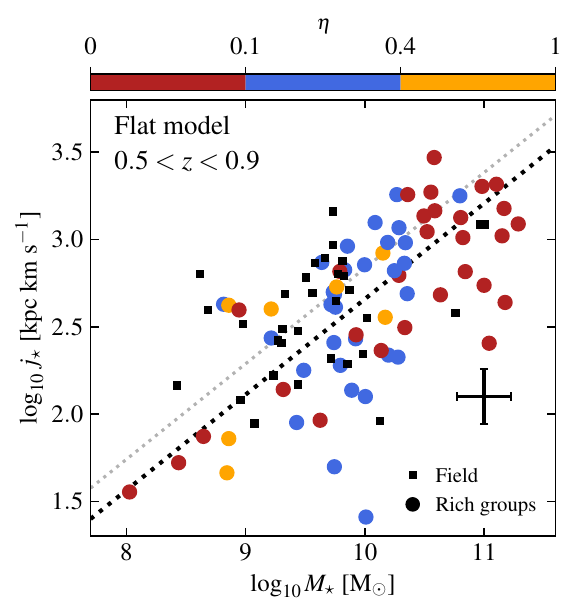}
		\caption{Fall relation colour coded according to $\eta$ (see Sect.\,\ref{sec:sample/environment}). This figure is similar to Fig.\,\ref{fig:analysis/environment/Fall relation eta} but uses the flat model rotation curve instead.}
		\label{fig:appendix/plots/Fall relation eta flat}
	\end{figure}
	
	\begin{figure}
		\centering
		\includegraphics[scale=0.9]{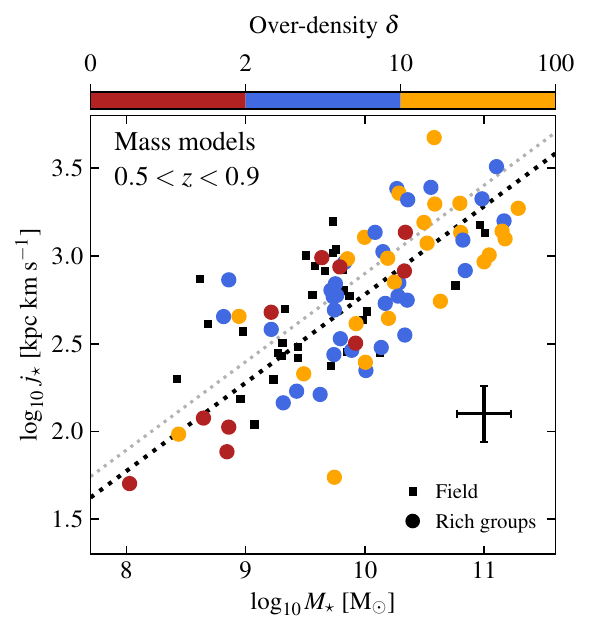}
		\caption{Fall relation colour coded according to the VMC method over-density estimate $\delta$. Only galaxies located in rich groups with more than ten members are shown as circles (black squares correspond to field galaxies). The mass model rotation curve is used with the redshift cut ($0.5 < z < 0.9$) applied. As an indication, the best-fit linear relations (with only the redshift cut applied, see also Table\,\ref{tab:fit_Fall_relation_structure_richness}) for the field and rich groups sub-samples with the stellar mass cut ($M_\star < \SI{e10}{\Msun}$) applied are shown as grey and black plain lines, respectively, and without the mass cut applied as grey and black dotted lines, respectively. The typical uncertainty is shown as the black error-bar.}
		\label{fig:appendix/plots/Fall relation Voronoi Brian}
	\end{figure}
	
	\begin{figure}
		\centering
		\includegraphics[scale=0.9]{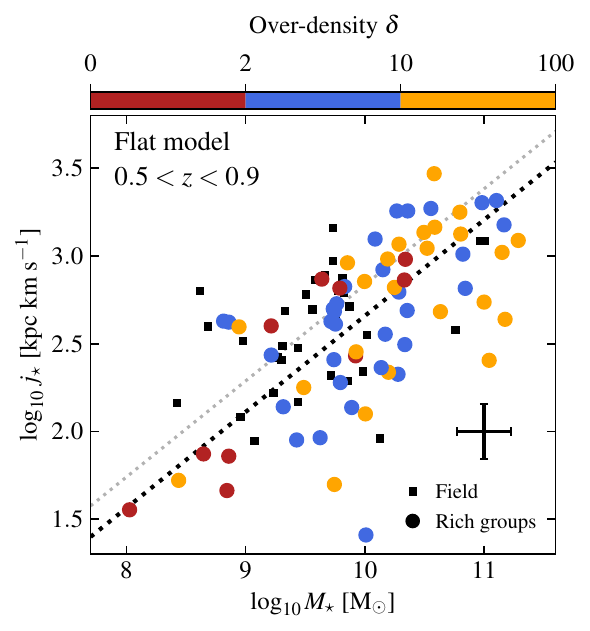}
		\caption{Fall relation colour coded according to the VMC method over-density estimate $\delta$. This figure is similar to Fig.\,\ref{fig:appendix/plots/Fall relation Voronoi Brian} but uses a flat model rotation curve instead.}
		\label{fig:appendix/plots/Fall relation Voronoi Brian flat}
	\end{figure}
	
	\begin{figure}
		\centering
		\includegraphics[scale=0.9]{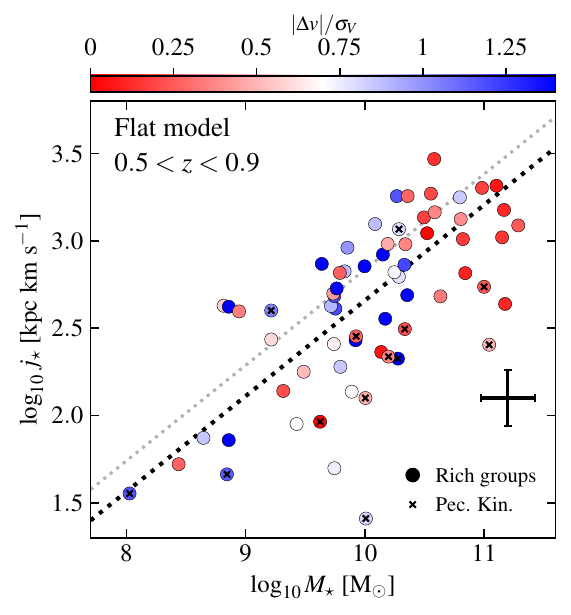}
		\caption{Fall relation colour coded according to the galaxies' line-of-sight systemic velocity $|\Delta v|$ normalised by the velocity dispersion of their host group $\sigma_V$. This figure is similar to Fig.\,\ref{fig:analysis/environment/Fall relation dV_sigma} but uses a flat rotation curve model instead}
		\label{fig:appendix/plots/Fall relation dV_sigma flat}
	\end{figure}
	
	\begin{figure}
		\centering
		\includegraphics[scale=0.9]{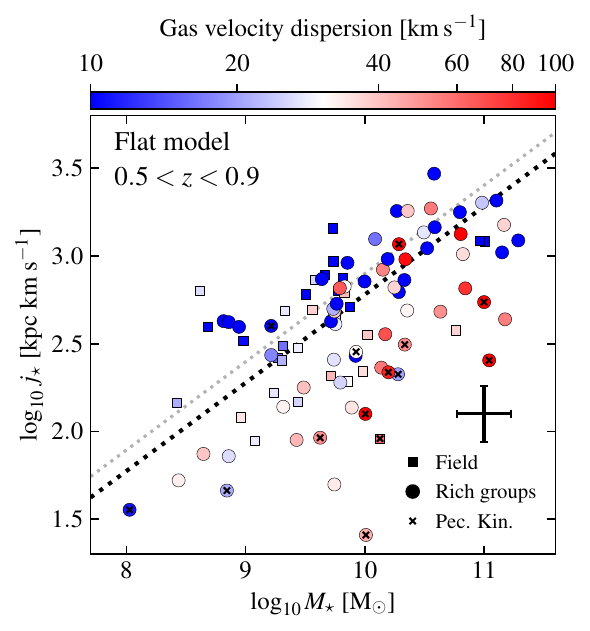}
		\caption{Fall relation colour coded according to the median velocity dispersion of the ionised gas component in the galaxies. This figure is similar to Fig.\,\ref{fig:analysis/environment/Fall relation gas dispersion} but uses a flat rotation curve model instead}
		\label{fig:appendix/plots/Fall relation gas dispersion flat}
	\end{figure}
	
%	\begin{figure*}
%		\centering
%		\includegraphics[scale=0.72]{momentum_mass_in_R22_normR22_vs_eta_flat}
%		\caption{Fall relation colour coded according to $\eta$ (left-hand plot), the galaxies' line-of-sight velocity $\Delta v$ normalised by the velocity dispersion of their host group $\sigma_V$ (middle plot), and the galaxies' distance to the centre of their host group normalised by the virial radius $R_{200}$ of the group (right-hand plot). This figure is similar to Fig.\,\ref{fig:analysis/environment/Fall relation eta} but uses the flat model rotation curve instead. The best-fit lines and the typical uncertainty correspond to those of the flat model rotation curve as well.}
%		\label{fig:appendix/plots/Fall relation eta flat}
%	\end{figure*}

\begin{table}
		\centering
		\caption{Median, 16th, and 84th percentiles estimates of the depletion of specific stellar angular momentum$^{\dagger\dagger}$ at the high-mass end when using the groups' richness as density estimate.}
		\resizebox{0.48\textwidth}{!}{%
		\begin{tabular}{c|lccp{17pt}l}
		\hline\\[-9pt]
		& Sub-sample & Stellar mass bin & $\Delta \beta$ & \multicolumn{2}{c}{16th, 84th}\\ 
		& & [$\log_{10} \unit{\Msun}$] & [\unit{\dex}] & \multicolumn{2}{c}{percentiles [\unit{\dex}]}\\
		& (1) & (2)        & (3) & \multicolumn{2}{c}{(4)}\\
		\hline
		\hline
		\multirow{6}{*}{\rotatebox[origin=c]{90}{Mass model}} & \multirow{3}{*}{Field$^\dagger$} & \hspace{5pt}$9.5$, $10.5$ & $-0.07$ & $-0.49,$ & \hphantom{- }$0.15$\\
		& & $10$, $11$ & $-0.74$ & $-0.83,$ & $-0.57$\\
		& & $10.5$, $11.5$ & $-0.78$ & $-0.85,$ & $-0.73$\\
		\cline{2-6}\\[-9pt]
		& \multirow{3}{*}{\bf{Large}} & \bf\bm{$9.5$}, \bm{$10.5$} & \bm{$-0.17$} & \bm{$-0.47,$} & \hphantom{- }\bm{$0.13$}\\
		& & \bf\bm{$10$}, \bm{$11$} & \bm{$-0.32$} & \bm{$-0.59,$} & \hphantom{- }\bm{$0.02$}\\
		& & \bf\bm{$10.5$}, \bm{$11.5$} & \bm{$-0.53$} & \bm{$-0.77,$} & \bm{$-0.23$}\\
		\hline
		\hline
		\multirow{6}{*}{\rotatebox[origin=c]{90}{Flat model}} & \multirow{3}{*}{Field$^\dagger$} & \hspace{5pt}$9.5$, $10.5$ & $-0.11$ & $-0.62,$ & \hphantom{- }$0.14$ \\
		& & $10$, $11$ & $-1.04$ & $-1.22,$ & $-0.71$\\
		& & $10.5$, $11.5$ & $-0.91$ & $-1.01,$ & $-0.87$\\
		\cline{2-6}\\[-9pt]
		& \multirow{3}{*}{\bf{Large}} & \bf\hspace{5pt}\bm{$9.5$}, \bm{$10.5$} & \bm{$-0.22$} & \bm{$-0.65,$} & \hphantom{- }\bm{$0.07$} \\
		& & \bf\bm{$10$}, \bm{$11$} & \bm{$-0.35$} & \bm{$-0.76,$} & \bm{$-0.08$}\\
		& & \bf\bm{$10.5$}, \bm{$11.5$} & \bm{$-0.63$} & \bm{$-1.01,$} & \bm{$-0.29$}\\
		\hline
		\end{tabular}}
		\label{tab:depletion_momentum_richness}\vspace{5pt}
			{\small\raggedright {\bf Notes:} (1) Sub-sample (i.e. galaxies in the field or in rich groups), (2) minimum and maximum bounds of the stellar mass bin, (3) median depletion of specific stellar angular momentum in the given mass bin with respect to the best-fit line obtained when fitting low-mass ($M_\star < \SI{e10}{\Msun}$) galaxies in the same sub-sample, and (4) 16th and 84th percentiles for the considered mass bin.\\ $^\dagger$ Values for the field sub-sample are given as an indication but there are too few galaxies in this sub-sample for all these stellar mass ranges to derive robust constraints.\\
$^{\dagger\dagger}$ The depletion is evaluated as the best-fit model (with mass and redshift cuts applied) minus the galaxies' specific stellar angular momentum. The same slope, corresponding to mass and redshift cuts applied (see Table\,\ref{tab:fit_Fall_relation_structure_richness}), was used for both sub-samples.\par}
	\end{table}	
	
	\clearpage
	\onecolumn
	\section{Examples of various types of morpho-dynamical models}
	
	\begin{figure*}[htbp]
		\centering
		\includegraphics[scale=0.5]{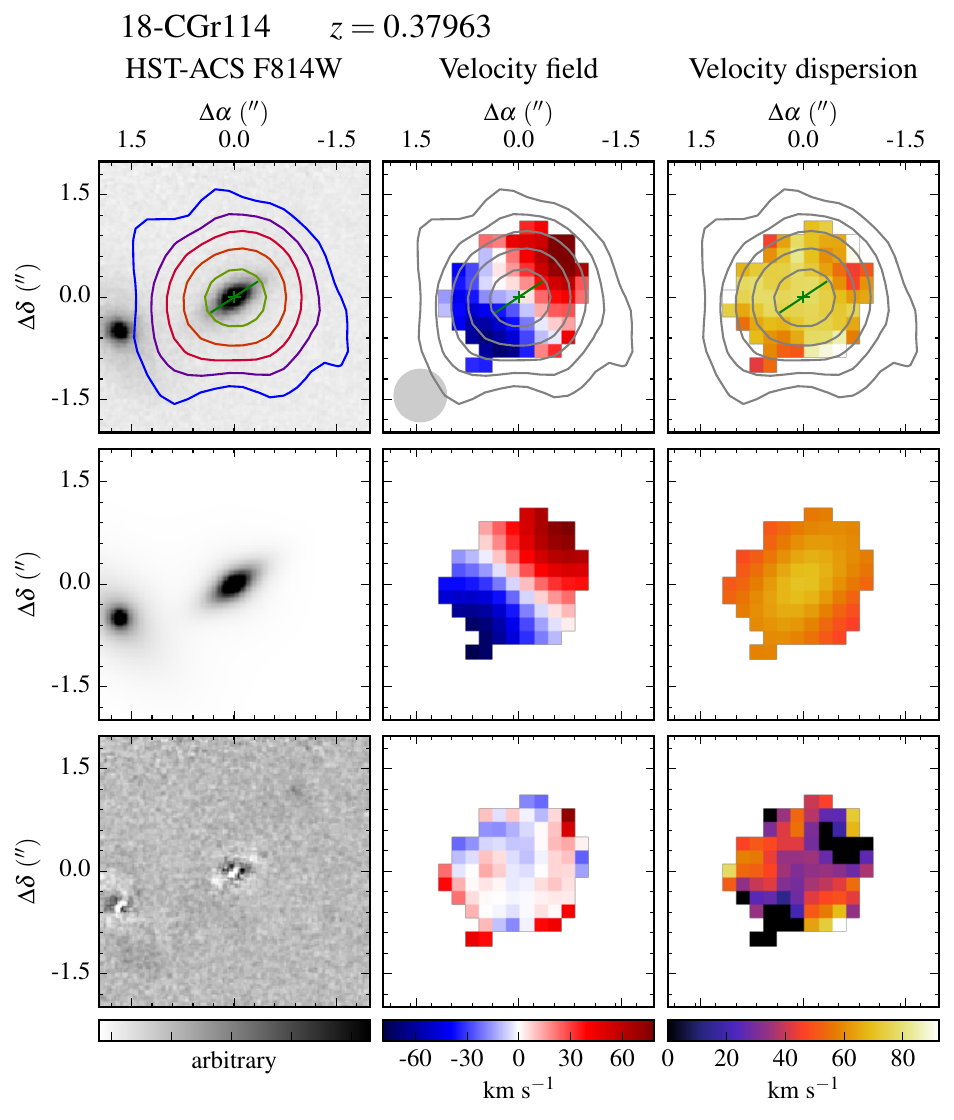}
		\includegraphics[scale=0.5]{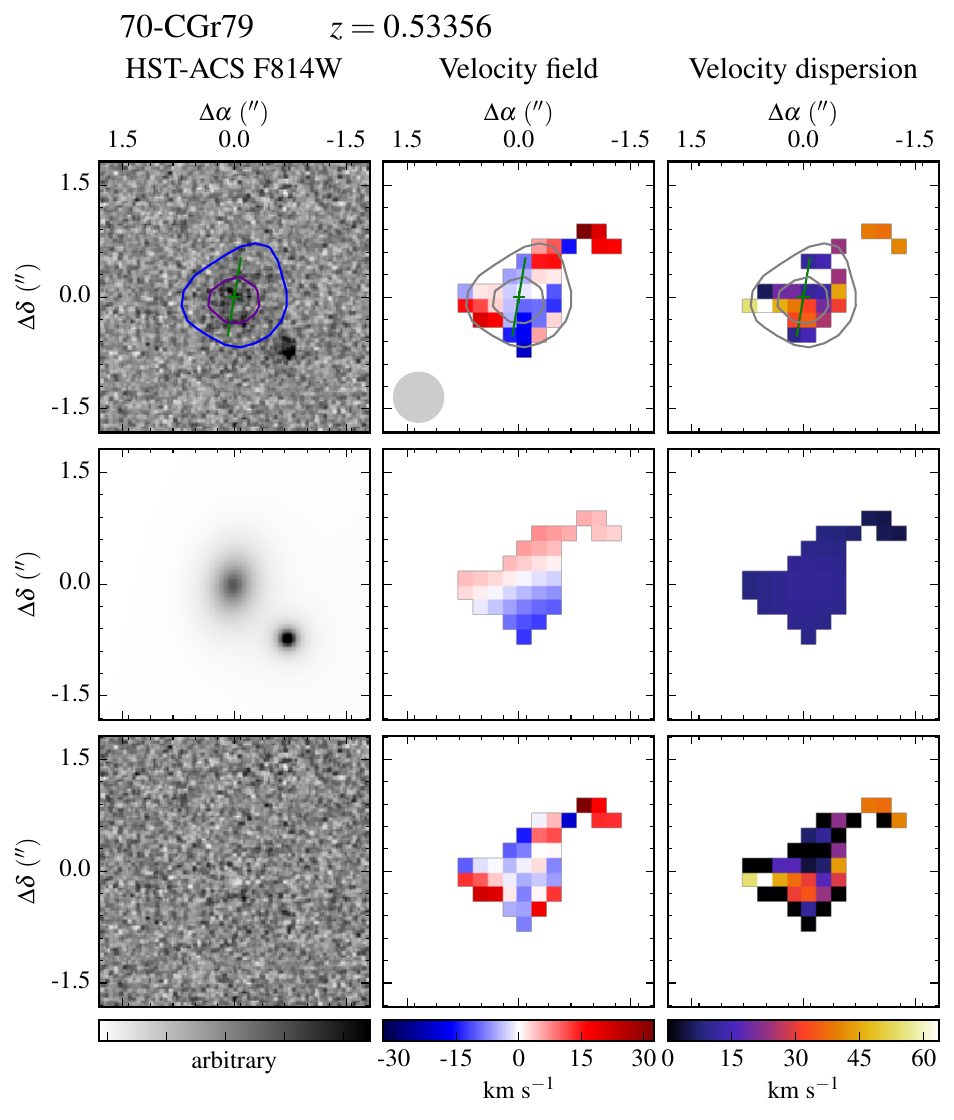}
		\includegraphics[scale=0.5]{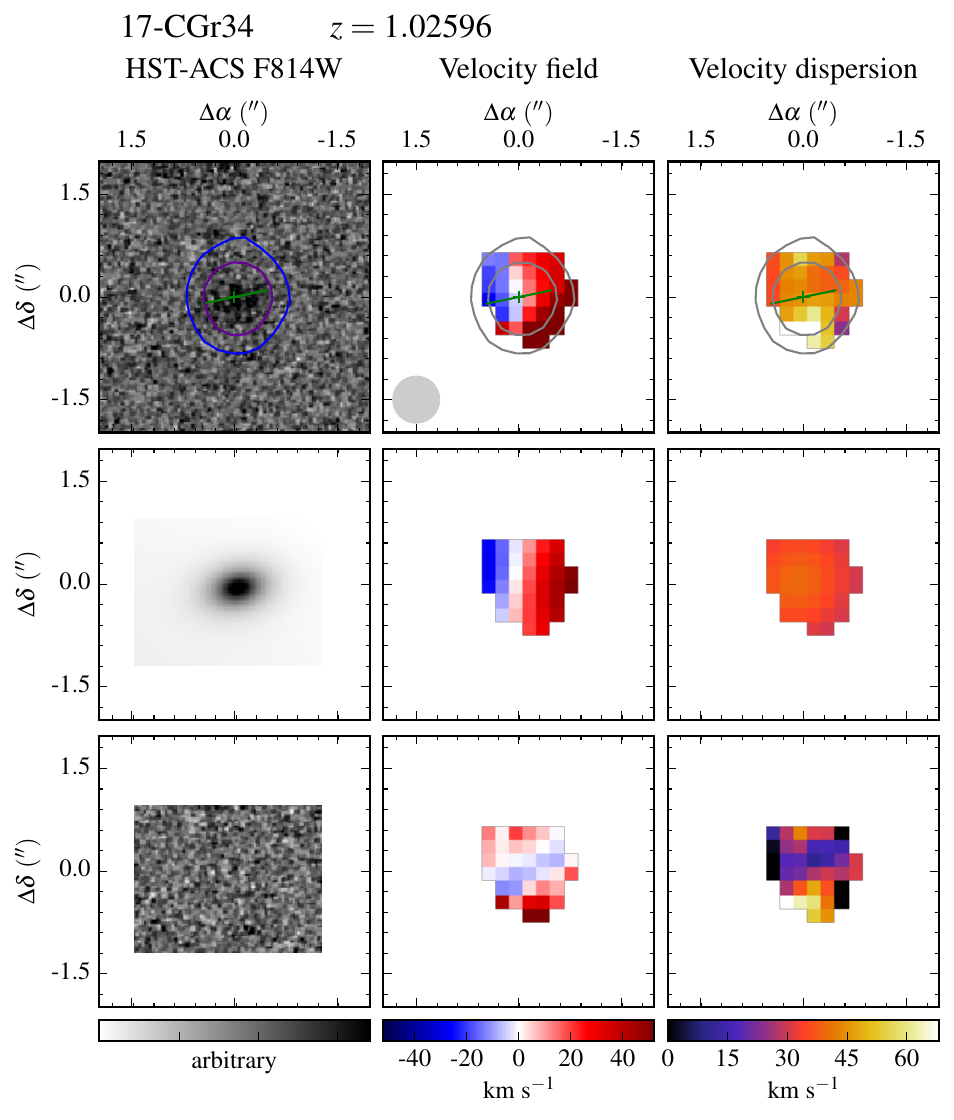}
		\includegraphics[scale=0.5]{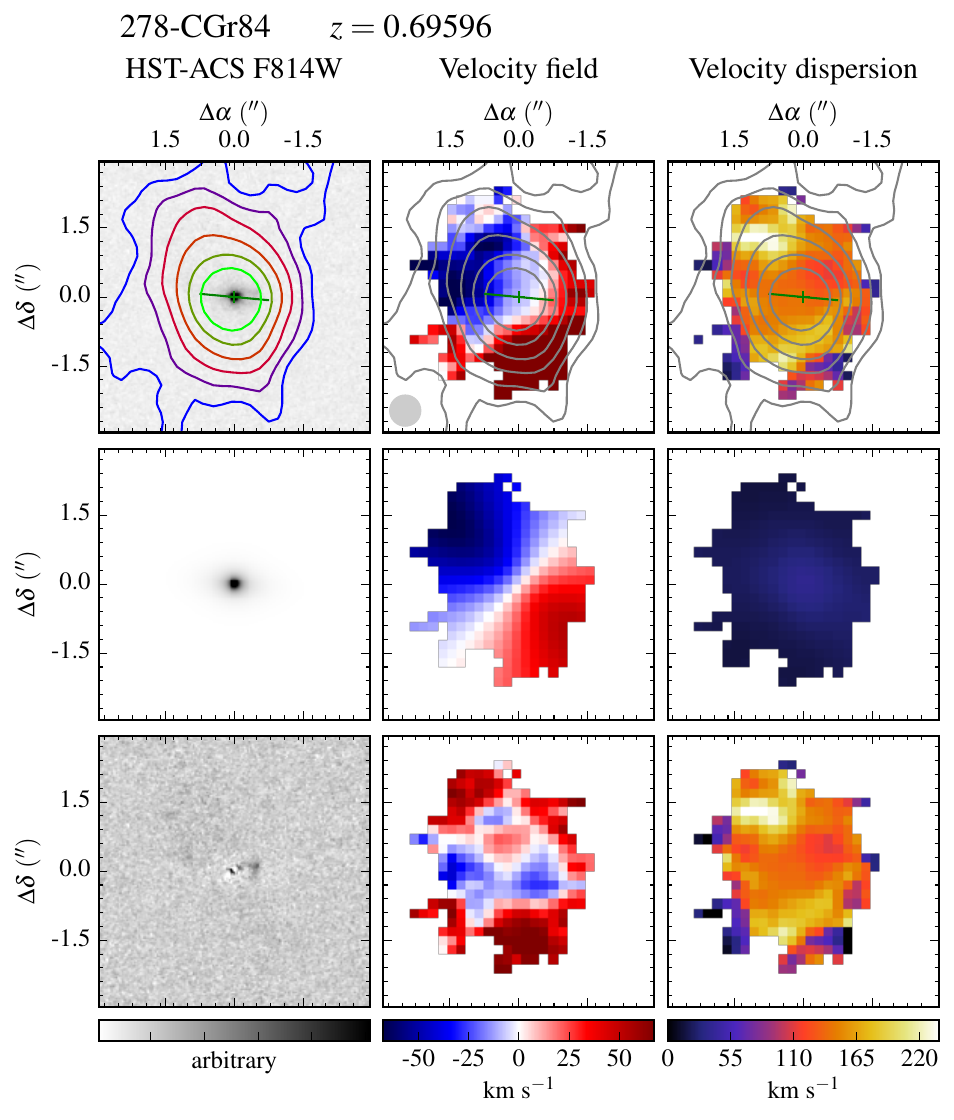}
		\caption{Examples of morpho-dynamical models obtained in \citet{mercier_scaling_2022} and used in this analysis. For each sub-figure and from top to bottom we have the \HST{} image, the \Galfit{} model, and the residuals in the leftmost column, the velocity field map extracted with \Camel{}, the best-fit \Mocking{} velocity field, and the residuals in the middle column, and the raw velocity dispersion map extracted with \Camel{}, the beam-smearing model from \Mocking{}, and the beam-smearing corrected velocity dispersion map from \Mocking{} in the rightmost column. The top left sub-figure shows a galaxy with a seemingly correct dynamical model, the top right one shows a low S/N galaxy without any velocity gradient in its velocity field map that was flagged, the bottom left one shows a low S/N galaxy with a weak but non-negligible velocity gradient that was not flagged, and the bottom right one shows a massive galaxy with a kinematically disturbed velocity field that was flagged as well.}
		\label{fig:sample/selection/examples dynamical models}
	\end{figure*}
	
	\clearpage
	\twocolumn
	\section{Updates on the morphological models}
	\label{appendix:updates_morpho}
	
	For this analysis and after inspecting the \HST{} images, we decided to remodel the morphology for the 17 following galaxies: \textsc{85\_CGr23}, \textsc{82\_CGr30}, \textsc{97\_CGr32}, \textsc{35\_CGr35}, \textsc{74\_CGr35}, \textsc{127\_CGr35}, \textsc{94\_CGr51}, \textsc{36\_CGr61}, \textsc{39\_CGr61}, \textsc{52\_CGr61}, \textsc{121\_CGr61}, \textsc{41\_CGr79}, \textsc{23\_CGr84}, \textsc{122\_CGr84}, \textsc{79\_CGr114}, \textsc{31\_CGr172}, and \textsc{68\_CGr172} (see Table\,F.1 of \citealp{mercier_scaling_2022} or Epinat et al., in prep. for the galaxies' IDs). Among these galaxies, we included a logarithmic spiral pattern to the disk component\footnote{Available since version three of \Galfit{}} for the four following galaxies: \textsc{85\_CGr23}, \textsc{121\_CGr61}, \textsc{36\_CGr61}, and \textsc{79\_CGr114}. We did so because these galaxies showed strong spiral and bar features so that their previous disk morphological model used to fit the bar rather than the extended disk. Even if the modelling of the spiral arms and of the bar is not trivial and far from perfect, we did find more robust disk parameters once these features were included, especially for the disk axis ratio and PA. Overall, the remaining 13 galaxies had a centre that was slightly misaligned with the bulge component which had the effect to mildly underestimate its contribution. With the updated models, we now have better constraints on the galaxy's centre and on the parameters of the bulge component. Furthermore, after trying to redo their morphology without more success, we decided to remove the 22 following galaxies:
	\begin{enumerate*}[label=(\roman*)]
		\item \textsc{12\_CGr61}, \textsc{24\_CGr32}, \textsc{26\_CGr61}, \textsc{29\_CGr23}, \textsc{54\_CGr79}, \textsc{76\_CGr30}, \textsc{76\_CGr84} \textsc{76\_CGr172}, \textsc{89\_CGr79}, \textsc{240\_CGr30}, and \textsc{240\_CGr84} because they are strongly bulge-dominated (probably elliptical) whose disks are too faint to be efficiently constrained, 
		\item \textsc{55\_CGr61}, \textsc{66\_CGr30}, \textsc{67\_CGr87}, \textsc{73\_CGr61}, \textsc{75\_CGr84}, \textsc{76\_CGr172}, \textsc{79\_CGr61}, \textsc{82\_CGr87}, and \textsc{97\_CGr34} because their disks are detected but with too low S/N to be properly constrained, and
		\item \textsc{79\_CGr30} and \textsc{84\_CGr35} because they both show complex edge-on morphologies that do not match their velocity fields. Indeed, \textsc{79\_CGr30} looks like two edge-on galaxies whose major axes are aligned but with a velocity field that spans the two objects and that shows undisturbed rotation with a PA rotated by $\SI{90}{\degree}$ with respect to the stellar components. Similarly, galaxy \textsc{84\_CGr35} seems to show multiple diffuse components that might be the result of a past merger or a stripping event even though its velocity field also shows an extended rotation which is not aligned with the brightest component visible in the \HST{} image. Therefore, the morphological and mass models for these two objects would be too uncertain to compute their angular momentum.	
	\end{enumerate*}
	
	\section{Deriving the angular momentum}
	\label{appendix:angular momentum}
	
	Our goal is to derive the angular momentum of the stars in the \MAGIC{} survey. We can split distribution of stars into two main components: the stellar disk and the stellar bulge. Among the two, the disk is the simplest to derive its angular momentum because of the assumptions that are usually made on its geometry and on its kinematics. Besides, bulges are usually thought of as stellar systems dynamically supported by random motions. In our case, because we model the bulge component as a spherical system, this means that its angular momentum should be nearly null. Thus, we will make the assumption that the bulge component does not posses any angular momentum and we will focus on deriving that of the disk component only.
	
	\subsection{General derivation}
	\label{appendix:angular momentum/disk}
	
	The angular momentum of a system is defined from its 3D mass distribution $\rho (\bm r)$ and 3D velocity field $\bm V (\bm r)$ as
	
	\begin{equation}
		\bm J = \int_{\mathcal{V}} d\tau~\rho(\vec r)~\bm r  \times \vec V (\vec r),
		\label{eq:appendix/angular momentum/disk/definition}
	\end{equation}
	where $\mathcal{V}$ represents the volume over which the integral is computed. In cylindrical coordinates, the cross product writes
	
	\begin{equation}
		\vec r \times \vec V (\vec r) = R V_\theta~\vec{\hat z} + (z V_R - R V_z )~\vec{\hat \theta} - z V_\theta~\vec{\hat R},
		\label{eq:appendix/angular momentum/disk/cross_product}
	\end{equation}
	where $\vec V (\vec r) = V_R~\vec{\hat R} + V_\theta~\vec{\hat \theta} + V_z~\vec{\hat z}$\footnote{In the following, when the velocity field is fully described by $V_\theta$, we will refer to it as the galaxy rotation curve.}. Assuming we want to derive the angular momentum of a disk-like distribution where only the tangential component of the velocity field is non-zero and does not depend on $z$, that is $V_R = V_z = 0$ and $V_\theta = V_\theta (R, \theta)$, then the angular momentum simplifies to
	
	\begin{equation}
		\vec J = \int_{\mathcal{V}} d\theta~dR~dz~\rho(\vec r)~\left [ R^2 V_\theta (R, \theta)~\vec{\hat z} - R z V_\theta (R, \theta)~\vec{\hat R} \right ].
		\label{eq:appendix/angular momentum/disk/general}
	\end{equation}
	
	\subsection{Angular momentum of thick disks}
	\label{appendix:angular momentum/thick disks}
	
	The angular momentum of a disk galaxy with a non-zero thickness profile $f(z)$ is similar to that of a razor-thin disk galaxy with the same surface mass density $\Sigma_M (R)$ if the three following criteria are met:
	
	\begin{enumerate}[label=(\roman*)]
		\item $\rho (\vec r) = \Sigma_{\rm{M}} (R, \theta) f(z)$,
		\item $\int_R dz~f(z) = 1$,
		\item $f(z) = f(-z)$.
	\end{enumerate}
	
	Criterion (i) means that it is possible to separate the surface mass density profile from the thickness profile. For instance, this is the case for a double exponential profile where $f(z)~=~\exp \left \lbrace- | z/h_z | \right \rbrace / h_z$. Among the three conditions, criterion (ii) is always met since it is required in order to recover the surface mass density when integrating the 3D mass distribution along the vertical axis with respect to the disk plane, that is $\Sigma_{\rm{M}} (R, \theta) = \int_{\mathbb{R}} dz~\rho(\vec r)$. Finally, criterion (iii) is usually assumed to simplify the shape of the vertical profile. If these three conditions are met, one can see that the rightmost integral in Eq.\,\ref{eq:appendix/angular momentum/disk/general} vanishes when integrating along the vertical axis, whereas only the $R$ and $\theta$ integrations remain in the leftmost one. Thus, we get an angular momentum integral similar to that of a razor-thin disk\footnote{For a razor-thin disk we have $f(z) = \delta(z)$, with $\delta$ a Dirac distribution, so criteria (i) to (iii) are also met.}
	
	\begin{equation}
		\vec J = \int_{\mathcal{S}} d\theta~dR~R^2~\Sigma_{\rm{M}} (R, \theta) V_\theta(R, \theta)~\vec{\hat z},
		\label{eq:appendix/angular momentum/thick disks/disk_thin}
	\end{equation}
	where $\mathcal{S}$ represents the surface over which the $R$ and $\theta$ integrations are carried out. It is common practice to normalise the angular momentum by the total stellar mass to define a specific angular momentum $j = | \vec J | / M_{\star}$ \citep[e.g.][]{Fall1983, romanowsky_angular_2012, bouche_muse_2021} but it is also possible to normalise it by the mass of the disk only. For instance, if we normalise the angular momentum measured over some surface $\mathcal{S}$ by the total disk mass, we get
	
	\begin{equation}
		j = \frac{\int_{\mathcal{S}} d\theta~dR~R^2~\Sigma_{\rm{M}} (R, \theta) V_\theta(R, \theta)}{\int_{[0, 2\pi) \times \mathbb{R}_{+}} d\theta~dR~R~\Sigma_{\rm{M}} (R, \theta)},
		\label{eq:appendix/angular momentum/thick disks/specific_angular_momentum_general}
	\end{equation}
	or in the case where there is no dependence on $\theta$ for both the surface mass density and the velocity:
	
	\begin{equation}
		j = \frac{\int_{R} dR~R^2~\Sigma_{\rm{M}} (R) V_\theta(R)}{\int_{\mathbb{R}_{+}} dR~R~\Sigma_{\rm{M}} (R)}.
		\label{eq:appendix/angular momentum/thick disks/specific_angular_momentum_simplified}
	\end{equation}
	
	\subsection{Angular momentum for special rotation curves}
	
	In most general terms, if one wants to compute the specific angular momentum for any surface mass density and velocity profile, one needs to numerically integrate Eq.\,\ref{eq:appendix/angular momentum/thick disks/specific_angular_momentum_general} or \ref{eq:appendix/angular momentum/thick disks/specific_angular_momentum_simplified} depending on whether there is a $\theta$ dependence or not. However, there exist for a few specific velocity profiles simplified formulae which can be computed either numerically or analytically in certain cases. We present in the following simplified expressions for some rotation curves, assuming no $\theta$ dependence on both the surface mass density and the rotation curve.
	
	\subsubsection{Constant rotation curve}
	\label{appendix:angular momentum/special/constant curve}
	
	If one assumes a constant rotation curve $V_\theta (R) = V_\theta$, then the angular momentum up to radius $r$ is given by
	
	\begin{equation}
		j(r) = V_\theta \times M_2 \lbrace \Sigma_{\rm{M}} \rbrace_0^r / M_1 \lbrace \Sigma_{\rm{M}} \rbrace_0^\infty,
		\label{eq:appendix/angular momentum/special/constant curve/specific_angular_momentum}
	\end{equation}
	where $M_k \lbrace \Sigma_{\rm{M}} \rbrace_\alpha^\beta$ is the $k^{\rm{th}}$ radial moment of the surface brightness distribution $\Sigma_{\rm{M}}$ integrated between $\alpha$ and $\beta$, that is
	
	\begin{equation}
		M_k \lbrace \Sigma_{\rm{M}} \rbrace_\alpha^\beta = \int_\alpha^\beta dR~R^k~\Sigma_{\rm{M}} (R).
		\label{eq:appendix/angular momentum/special/constant curve/radial_momentum_definition}
	\end{equation}
	
	\subsubsection{Flat model}
	\label{appendix:angular momentum/special/flat model}
	
	If one assumes the rotation curve to be a flat model as in \citet{Abril-Melgarejo2021}, that is described by the following formula:
	
	\begin{equation}
		V_\theta (R) = V_{\rm{t}} \times
		\begin{cases}
			R / R_{\rm{t}}, & \mbox{if } R < R_{\rm{t}} \\
			1, & \mbox{otherwise}
		\end{cases}
		\label{eq:appendix/angular momentum/special/flat model/flat_model}
	\end{equation}
	then the angular momentum is given by
	
	\begin{equation}
		j(r) = \frac{V_{\rm{t}}}{R_{\rm{t}} M_1 \lbrace \Sigma_{\rm{M}} \rbrace_0^\infty} \times
		\begin{cases}
			M_3 \lbrace \Sigma_{\rm{M}} \rbrace_0^R, & \mbox{if } R < R_{\rm{t}} \\
			M_3 \lbrace \Sigma_{\rm{M}} \rbrace_0^{R_{\rm{t}}} + R_{\rm{t}} M_2 \lbrace \Sigma \rbrace_{R_{\rm{t}}}^r, & \mbox{otherwise}
		\end{cases}
		\label{eq:appendix/angular momentum/special/flat model/specific_angular_momentum}
	\end{equation}
	
	\subsection{Angular momentum for a Sérsic profile}	
	\label{appendix:angular momentum/special/sersic}
	
	If the surface mass density can be further described by a Sérsic profile \citep{Sersic_model} with Sérsic index $n$, effective radius $R_{\textit{eff}}$ and central surface mass density $\Sigma_{{\rm{M}}, 0}$, then we have
	
	\begin{equation}
		\vec J = \Sigma_{M, 0} \int_R dR~R^2 e^{-b_n (R/R_{\textit{eff}})^{1/n}} \int_\theta d\theta~V_\theta (R, \theta)~\vec{\hat z},
	\end{equation}
	where $b_n$ is the solution of the equation $\Gamma \left (2n \right )~=~2\gamma\left (2n, b_n \right )$ \citep{Graham2005}, with $\gamma$ and $\Gamma$ the lower incomplete and complete gamma functions, respectively. Without any $\theta$ dependence, the specific angular momentum writes
	
	\begin{equation}
		j = \frac{2~b_n^{2n}}{(2n)!} \int_R dR~(R/R_{\textit{eff}})^2~V_\theta (R)~e^{-b_n (R/R_{\textit{eff}})^{1/n}},
		\label{eq:appendix/angular momentum/special/sersic/specific_angular_momentum}
	\end{equation}
	where the normalisation factor comes from the analytical expression for the total mass of a Sérsic profile \citep[e.g.][]{mercier_scaling_2022}. In the case of a flat rotation curve as in Eq.\,\ref{eq:appendix/angular momentum/special/flat model/flat_model}, it is possible to compute analytically Eq.\,\ref{eq:appendix/angular momentum/special/constant curve/radial_momentum_definition} and therefore Eq.\,\ref{eq:appendix/angular momentum/special/flat model/specific_angular_momentum}. Indeed, we have
	
	\begin{equation}
		\begin{split}
			M_k \lbrace \Sigma_{\rm{M}} \rbrace_\alpha^\beta = \frac{n~\Sigma_{{\rm{M}}, 0}~R_{\textit{eff}}^{k+1}}{b_n^{n(k+1)}} \left [ \gamma \left ( n(k+1), b_n \left ( \frac{\beta}{R_{\textit{eff}}} \right )^{1/n} \right ) \right. \\
		 \left. - \gamma \left ( n(k+1), b_n \left (\frac{\alpha}{R_{\textit{eff}}} \right )^{1/n} \right ) \right ]
		\end{split}.
		\label{eq:appendix/angular momentum/special/sersic/radial_moment_sersic_flat}
	\end{equation}
	
	\subsection{Recovering the RF12 approximation}
	
	The \citet{romanowsky_angular_2012} approximation, hereafter referred as RF12, assumes an exponential disk with a constant rotation curve and computes the angular momentum at infinity. Thus, we can use Eq.\,\ref{eq:appendix/angular momentum/special/flat model/specific_angular_momentum} and \ref{eq:appendix/angular momentum/special/sersic/radial_moment_sersic_flat} to check whether we recover it. First, we can see that for a flat model and in the limit where the rotation curve becomes constant, that is $R_{\rm{t}} \rightarrow 0$, Eq.\,\ref{eq:appendix/angular momentum/special/flat model/specific_angular_momentum} reduces to \ref{eq:appendix/angular momentum/special/constant curve/specific_angular_momentum}. Furthermore, evaluated at infinity and for an exponential disk model (i.e. $n=1$) Eq.\,\ref{eq:appendix/angular momentum/special/sersic/radial_moment_sersic_flat} simplifies to
	
	\begin{equation}
		M_k \lbrace \Sigma_{\rm{M}} \rbrace_0^{\infty} = \Sigma_{\rm{M}, 0} R_{\rm{d}}^{k+1} \Gamma (k+1),
	\end{equation}
	where $\Gamma$ is the complete gamma function and $R_{\rm{d}} = R_{\textit{eff}, \rm{d}} / b_1$ is the disk scale length, with $b_1 \approx 1.6783$. Thus, given Eq.\,\ref{eq:appendix/angular momentum/special/constant curve/specific_angular_momentum}, we recover the \citet{romanowsky_angular_2012} approximation
	
	\begin{equation}
		j = 2 V_\theta R_{\rm{d}}.
	\end{equation}
	
	\subsection{Angular momentum for an exponential disk in equilibrium}
	\label{appendix:angular momentum/special/disk}
	
	It can be interesting to derive the angular momentum for an exponential disk model that is stable against its own gravity, that is without any DM halo or bulge component. We consider the case where the normalisation is taken at the same radius as the angular momentum. Given the expression of the integrated mass and circular velocity for such a model (see Eqs.\,C.1, D.8, and D.9 of \citet{mercier_scaling_2022}, its integral writes
	
	\begin{equation}
		j (R) = \frac{\sqrt{\pi G R_{\rm{d}} \Sigma_{\rm{M, d}} (0)}}{2 \pi R_{\rm{d}}^3 \gamma(2, R/R_{\rm{d}})} \int_0^R dr\,r^3 e^{-r/R_{\rm{d}}} f \left ( \frac{r}{2 R_{\rm{d}}} \right ),
	\end{equation}
	where $G$ is the gravitational constant, $R_{\rm{d}}$ is the disk scale length, $\Sigma_{\rm{M, d}} (0)$ is the central surface mass density of the disk, $\gamma$ is the lower incomplete gamma function, and $f$ is a function of Bessel functions defined in Appendix\,D of \citet{mercier_scaling_2022}. If we define $y = x / R_{\rm{d}}$, then it simplifies to
	
	\begin{equation}
		j (R) = \frac{\sqrt{G M_{\rm{d}} (R) R_{\rm{d}}}}{2\sqrt{2} \pi \gamma^2(2, R/R_{\rm{d}})} \times \int_0^{R/R_{\rm{d}}} dy\,y^3 e^{-y} f(y/2),
		\label{eq:appendix/angular momentum/special/disk/simplied}
	\end{equation}
	where $M_{\rm{d}} (R)$ is the mass of the disk within the radius $R$. At $R_{22} = 2.2 R_{\rm{d}}$, the radius of maximum velocity for the disk, both $\gamma$ and the integral on the right-hand side become constants and Eq.\,\ref{eq:appendix/angular momentum/special/disk/simplied} further reduces to
	
	\begin{equation}
		j (R_{22})\,[\rm{kpc\,km\,s^{-1}}] \approx 3.373 \times 10^{-4} \times \sqrt{M_{\rm{d}} (R_{22}) R_{\rm{d}}},
		\label{eq:appendix/angular momentum/special/disk/R22}
	\end{equation}
	where $M_{\rm{d}} (R_{22})$ is in \unit{\Msun} and $R_{\rm{d}}$ in kpc.

	\clearpage
	\onecolumn
	\section{Catalogue of the main sample}%
	
	\begin{table*}[!htbp]
		  \centering
		  \caption{Column description of the catalogue used in this analysis.}
		  
	\begin{tabular}{lll}
		\hline
		\hline
		No.& Title & Description \\
		\hline\\
		1 & ID & MUSE galaxy ID in the form X-CGrY, where X refers to the galaxy identification \\ 
		 & & number within the field targeting COSMOS group CGrY \\
		2 & z & Spectroscopic systemic redshift derived from the kinematics modelling \\
		3 & RA & J2000 Right Ascension of morphological centre in decimal degrees \\
		4 & Dec & J2000 Declination of morphological centre in decimal degrees \\
		5 & N & Number of galaxies in the host structure. Field galaxies have a value of -99. \\ 
		6 & R/R200 & Ratio of the projected distance of a galaxy to its host group's centre with the radius of the \\
		& & group $R_{200}$ (see Sect.\,\ref{sec:sample/environment}). Field galaxies have a value of -99. \\
		7 & dV/sigma & Ratio of the systemic velocity of a galaxy along the line-of-sight with its host group's \\
		& & velocity dispersion (see Sect.\,\ref{sec:sample/environment}). Field galaxies have a value of -99. \\
		8 & Eta & Environmental tracer defined in Sect.\,\ref{sec:sample/environment}, combining parameters 6 and 7. Field galaxies have \\
		& & a value of -99. \\
		9 & log over-density & Logarithm of the over-density around the galaxies as measured by the VMC technique \\
		& &  (see the end of Sect.\,\ref{sec:sample/environment}) \\
		10 & log j\_HST\_MM & Logarithm of the specific stellar angular momentum (\unit{\kilo\pc \kilo\meter\per\second})  derived from HST images \\
		& &  using the mass model rotation curve \\
		11 & log j\_HST\_flat & Logarithm of the specific stellar angular momentum (\unit{\kilo\pc \kilo\meter\per\second})  derived from HST images  \\
		& &  using the flat model rotation curve \\
		12 & log j\_exp\_MM & Logarithm of the specific stellar angular momentum (\unit{\kilo\pc \kilo\meter\per\second})  derived by numerically \\
		& &  integrating Eq.\,\ref{eq:derivation/specific_angular_momentum_this_analysis} assuming an exponential disk mass distribution and a mass \\
		& & model rotation curve \\
		13 & log j\_exp\_flat & Logarithm of the specific stellar angular momentum (\unit{\kilo\pc \kilo\meter\per\second}) analytically derived  \\
		& & assuming an exponential disk mass distribution and a flat model rotation curve (see \\ 
		& &   Appendix\,\ref{appendix:angular momentum}) \\
		14 & Gas dispersion & Average velocity dispersion of the ionised gas (\unit{\kilo\meter\per\second}) measured in the velocity dispersion map \\
		& & of the \OII{} doublet after correcting for the LSF and beam-smearing \\
		15 & log M* & Logarithm of the stellar mass (\unit{M_\odot}) \\
		\hline\\
	\end{tabular}
	\label{tab:catalog_CDS}
	{\small\raggedright {\bf Notes:}  The sample used in this catalogue corresponds to the \Nselection{} galaxies selected in terms of their surface, S/N, and stellar disk inclination (see Sect.\,\ref{sec:sample/selection}). Complementary physical, morphological, and dynamical information can be found in the morpho-kinematics catalogue published in \citet[][catalogues can be cross-matched using the \textit{ID} field]{Mercier_2022_cat}. All logarithmic values are given as decimal logarithms.\par}
	\end{table*}
	
	\end{appendix}
\end{document}